\documentclass[journal=jctcce,manuscript=article]{achemso}
\usepackage[utf8]{inputenc}
\usepackage{textcomp}
\usepackage{graphicx}
\usepackage{mhchem}
\usepackage{siunitx}
\DeclareSIUnit\au{a.u.}
\usepackage{amsmath}
\usepackage{upgreek}
\usepackage{hyperref}
\usepackage{amssymb}
\usepackage{tabularx}
\usepackage{booktabs}
\usepackage{epstopdf}
\usepackage{diagbox}
\usepackage{threeparttable} 
\AppendGraphicsExtensions{.tiff}
\usepackage{comment}
\usepackage[shortlabels]{enumitem}

\DeclareMathOperator{\Tr}{Tr}

\newcommand{\EC}[0]{E_C}
\newcommand{\EX}[0]{E_X}
\newcommand{\XC}[0]{\chi_C}

\allowdisplaybreaks

\title{Spatial Signatures of Electron Correlation in Least-Squares Tensor Hyper-Contraction}

\author{Chao Yin}
\affiliation{Department of Chemistry, Southern Methodist University, Dallas, TX 75275, USA}

\author{Sara Beth Becker}
\affiliation{Department of Chemistry, Southern Methodist University, Dallas, TX 75275, USA}
\altaffiliation
{Present address: Department of Anthropology, Southern Methodist University, Dallas, TX 75275, USA}

\author{James H. Thorpe}
\affiliation{Department of Chemistry, Southern Methodist University, Dallas, TX 75275, USA}

\author{Devin A. Matthews}
\email{damatthews@smu.edu}
\affiliation{Department of Chemistry, Southern Methodist University, Dallas, TX 75275, USA}

\abbreviations{CCSD, LS-THC}

\begin{document}

\begin{tocentry}
\includegraphics[scale=1]{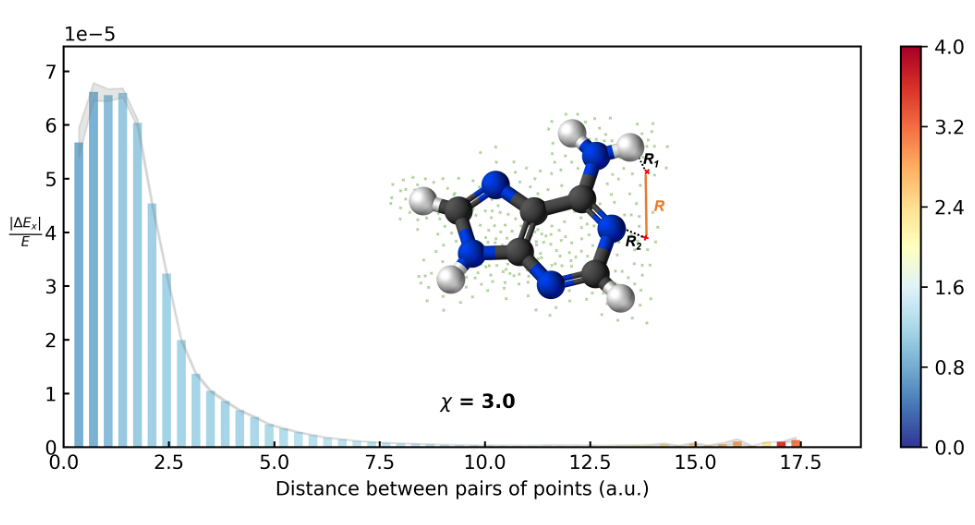}
\end{tocentry}

\begin{abstract}

Least Squares Tensor Hypercontraction (LS-THC) has received some attention in recent years as an approach to reduce the significant computational costs of wavefunction based methods in quantum chemistry. However, previous work has demonstrated that the LS-THC factorization performs disproportionately worse in the description of wavefunction components (e.g. cluster amplitudes $\hat{T}_2$) than Hamiltonian components (e.g. electron repulsion integrals $(pq|rs)$). This work develops novel theoretical methods to study the source of these errors in the context of the real-space $\hat{T}_2$ kernel, and reports, for the first time, the existence of a ``correlation feature'' in the errors of the LS-THC representation of the ``exchange-like'' correlation energy $\EX$ and $\hat{T}_2$ that is remarkably consistent across ten molecular species, three correlated wavefunctions, and four basis sets. This correlation feature portends the existence of a ``pair-point kernel'' missing in the usual LS-THC representation of the wavefunction, which critically depends upon pairs of grid points situated close to atoms and with inter-pair distances between one and two Bohr radii. These findings point the way for future LS-THC developments to address these shortcomings. 
\end{abstract}

\section{Introduction}
Recent years have seen a rapid growth in the use of tensor-factorization techniques to reduce the cost of deterministic, wavefunction-based quantum chemical simulations\cite{Whitten-JCP-1973,Dunlap-JCP-1979,Dunlap-PCCP-2000,Martin-CPL-1993,Vahtras-CPL-1993,Ren-NJP-2012,eichkorn-JPL-1995, eichkorn-TCA-1997, Weigend-PCCP-2002,Weigend-JCP-2009,Beebe-IQC-1977,Francesco-JCP-2007,Henrik-JCP-2003,Bell-MP-2010,Auer-JCP-2011,Auer-JCP-2013,Martinez-JCP-1992,Kokkila-JCTC-2015,Hohenstein-JCP-2012-1,Parrish-JCP-2012,Hohenstein-JCP-2012-2,Matthews-JCTC-2020,Matthews-JCP-2021,Schutski-JCP-2017,Gruneis-JCP-2017,Bartlett-JCP-2014,Lesiuk-JCC-2019,Lesiuk-JCTC-2020,Lesiuk-JCP-2022,Valeev-JCTC-2016,Valeev-JCTC-2021}. The fundamental driver of these developments is the fact that the cost of wavefunction construction and measurement of wavefunction properties -- most importantly the electronic energy -- invariably arises from computational operations involving two or more high dimensional tensors, the size and number of which grows exponentially with the accuracy demanded of the simulation. Techniques such as the Cholesky Decomposition \cite{Weigend-JCP-2009,Beebe-IQC-1977,Henrik-JCP-2003,Francesco-JCP-2007} reduce computational cost by factorizing the four-index tensors in the electronic Hamiltonian into products of three-index tensors, consequently lowering I/O costs, but not formal scaling. Going further, the Singular Value Decomposition\cite{Kinoshita-JCP-2003,Bartlett-JCP-2014,Bell-MP-2010,Parrish-JCP-2019,Lesiuk-JCC-2019} promises to reduce the cost of wavefunction-based methods by an order of magnitude or more by factorizing the doubles and/or triples amplitudes into products of lower-rank tensors, resulting in a reduction in the overall scaling of the method. An even more aggressive form of factorization, least-squares tensor hyper-contraction (LS-THC), shows promise in reducing computational cost further with low-rank tensors (now matrices) that are determined by fitting using simple, closed form equations\cite{Parrish-JCP-2012,Matthews-JCP-2021}. It is this latter factorization that is the focus on this work.

LS-THC is best understood in the context of the decomposition of the electron repulsion integrals (ERIs). In this context, the four-dimensional ERI tensor is represented as,
\begin{align}
    g^{pq}_{rs} = (pr|qs) \approx \sum_{RS} X_{p}^{R} X_{r}^{R} V_{RS} X_{q}^{S} X_{s}^{S}
\end{align}
where $g^{pq}_{rs}$ is an element of the ERI tensor $\mathbf{g}$, $\mathbf{X}$ is a collocation matrix that contains the weighted evaluation of molecular orbital $p$ at spatial grid point $R$ (typically $X_p^R = \phi_p(x_R)\omega_R^{1/4}$ where $\omega_R$ is the quadrature weight associated with grid point $x_R$), and $\mathbf{V}$ is the core matrix that fits $\mathbf{g}$ as interactions between spatial grid points. Importantly, the collocation matrices are trivially obtained a priori from the original orbitals and a set of spatial grid points, and so the core matrix may be simply constructed from a least squares fit of the original tensor and the collocation matrices,
\begin{align}
    V_{PQ} &= \sum_{RS}(S^{-1})_{PR}(S^{-1})_{SQ} \sum_{pqrs} X_p^{R} X_r^{R} X_q^{S} X_s^{S} g^{pq}_{rs} \\
    S_{RS} &= \sum_{pq} X^R_p X^R_q X^S_p X^S_q
\end{align}
Likewise, the well-known double amplitudes in the many-body representation of the wavefunction (either the coupled cluster or configuration interaction doubles can be used, although we focus on coupled cluster here) may be approximated as,
\begin{align}
    t^{ab}_{ij}  \approx \sum_{RS} X_{a}^{R} X_{i}^{R} T_{RS} X_{b}^{S} X_{j}^{S}
\end{align}
where $t^{ab}_{ij}$ are doubles amplitudes, $\mathbf{X}$ retains the same definitions as
before, and $\mathbf{T}$ is the core matrix that represents the wavefunction in the LS-THC grid space. The promise of LS-THC lays in the fact
that the $\mathbf{T}$ core matrix can be constructed without ever actually constructing the original amplitudes. In fact, by employing a clever combination of Density Fitting\cite{Whitten-JCP-1973,Dunlap-JCP-1979,Martin-CPL-1993,Dunlap-PCCP-2000,Vahtras-CPL-1993} (DF) and an inverse Laplace transform and quadrature of the orbital energy denominators\cite{Haser-JCP-1992,Constans-JCP-2000,Kats-PCCP-2008}, 
LS-THC representations of coupled cluster singles and doubles (CCSD) and similar wavefunctions never explicitly construct any rank-4 tensor. Instead, the integrals are
transformed into the MO basis via density fitting quantities (scaling as $\mathcal O(N^4)$, where $N$ is the number of molecular orbitals), fit into the LS-THC representation 
(scaling as $\mathcal O(N^4)$), used to construct compressed LS-THC residuals (scaling as $\mathcal O(N^4)$ after factorization), 
which are then fit into LS-THC $\mathbf{T}$ core matrix updates using the Laplace transform quadrature (scaling as $\mathcal O(N^3)$), resulting in a CCSD approximation that scales as $\mathcal O(N^4)$ overall\cite{Parrish-JCP-2012} (THC also scales as $\mathcal O(N^4)$ when using pre-selected projectors\cite{Hohenstein-JCP-2022} or in a fully non-linear fitting scheme\cite{Schutski-JCP-2017}).

However, this reduction in computational scaling depends upon one crucial assumption: the number of LS-THC grid points required to obtain 
the accuracy goals desired by the practitioner must scale as $\mathcal O(N)$. Further, the cost of a LS-THC calculation will still increase
at least cubically as the set of LS-THC grid points are expanded in the pursuit of a given accuracy goal, for a fixed number of orbitals. This would not be a problem if these 
methods converge quickly as the ratio of grid points to molecular orbitals ($\chi = n_g /N$) increases. Unfortunately, we and others have
demonstrated that this is decidedly not the case,\cite{lee-JCTC-2020,Matthews-JCP-2021} especially for basis sets larger than cc-pVDZ. As explained in that work, while the contraction of the LS-THC representation of the 
Coulomb-ordered ERIs and doubles amplitudes needs only a small number of grid points to obtain adequate accuracy, the contraction 
of exchange-ordered ERIs and doubles amplitudes (or, equivalently, contraction of Coulomb-ordered ERIs with exchange-ordered amplitudes) 
results in sizeable errors in valance exchange correlation energies that improve extremely slowly as the span of the LS-THC grid point 
basis increases. 

From whence does this pernicious feature of LS-THC arise? This is the question that this manuscript will attempt to answer, or, if not 
answer, at least explore. In Ref.~\citenum{Matthews-JCP-2021}, it was ascribed to the fact that the contraction of LS-THC factorized 
exchange-ordered ERIs and doubles amplitudes is fundamentally the contraction of a non-local, non-low-rank tensor and a 
local, low-rank tensor, and thus should not be expected to support rapid convergence with respect to a basis constructed of the local, 
spatial LS-THC grid points. There is, however, another perspective given by the recent work of Mardirossian, McClain, and Chan\cite{Mardirossian-JCP-2018},
that in the context of LS-THC suggests this ``fitting error'' may not be the fault of the representation of the integrals, but of the wavefunction. That work 
explored (continuous) spatial representations of the quantities discussed here and how they may be used to accelerate simulations. In that framework,
the molecular-orbital ERIs may be expressed as an integral of a two-center ERI kernel over the (real-valued) molecular orbitals,
\begin{align}\label{eq:eri}
    g^{pq}_{rs} =  \int \phi_p(r_1) \phi_r(r_1) \frac{1}{|r_1 - r_2|} \phi_q(r_2) \phi_s(r_2) dR
\end{align}
which makes clear the connection of LS-THC to a form of real-space quadrature of the above equation. The real-space wavefunction, however, is an integral
of a four-center amplitude kernel over the molecular orbitals,\cite{Mardirossian-JCP-2018}
\begin{align}\label{eq:t2-amplitude}
    t_{ij}^{ab} = \int \phi_a(r_1) \phi_i(r_1^{'}) \hat{t}(r_1, r_2, r_1^{'}, r_2^{'})  \phi_b(r_2) \phi_j(r_2^{'}) dR
\end{align}
Here the connection of LS-THC to a real-space quadrature is more tenuous, as the LS-THC core-matrix is only a two-centered entity. From this perspective the 
missing, four-center interactions of the amplitude kernel are the most likely cause of the error in LS-THC. This manuscript explores this point of view by utilizing a set of novel theoretical techniques that permit the separation of the correlation energy and correlated wavefunction into contributions arising from the usual, single-point grid basis, and those arising from an as-yet-unexplored basis that captures the interactions between pairs of spatial grid points, providing, for the first time, spatial resolution of the features of electron correlation responsible for the slowly decaying grid-point-dependence of LS-THC methods.

\section{Theoretical Methods}
\subsection{Notation}
Throughout this work we will employ the following notation:
\begin{itemize}
    \item{$\mu \rho \nu \sigma$: atomic orbitals (AOs), of number $M$.}
    \item{$pqrs$: molecular orbitals (MOs), of number $N$.}
    \item{$abcdef$: virtual MOs, of number $n_v$.} 
    \item{$ijklmn$: occupied MOs, of number $n_o$.}
    \item {$PQRS$:  LS-THC spatial grid points, of number $n_g$.}
    \item{$\chi$: the ratio of number of grid points to molecular orbitals, $\chi = n_g/N$.}
    \item{$\EC, \EX$: the ``Coulomb-like'' and ``exchange-like'' correlation energy contributions, respectively (vide infra). }
\end{itemize}
We avoid the use of Einstein notation when indices of tensors are specified. However, two or more adjacent tensors without specified indices indicates a tensor contraction cast into matrix multiplication form.

\subsection{Energy Analysis via Partitioning}
In this subsection we develop the methodology for investigating the influence of interactions between pairs of LS-THC grid points. Specifically, we are interested in how these pairs influence the portions of the valance correlation energy that arise from contractions of the ERIs with $\hat{T}_2$.  By  partitioning the contributions of a ``starting set'' of $N$ grid points, we may track the influence of the introduction of a new pair of points on the LS-THC representation of $\hat{T}_2$, and determine the influence of any given pairs of points we wish to study. We begin with a ``starting set'' of $N$ LS-THC grid points that define the single-point basis. The joint collocation matrix of this basis is given as
\begin{align}\label{eq:joint-cllocation-matrix}
    Y^P_{ai} \equiv X^P_a X^P_i
\end{align}
which can also be recognized as a Khatri-Rao product.\cite{KRP-wiley-2004,Bader-SIAM-2009} This joint collocation matrix projects an electron described by the distribution of orbitals $\phi_a$ and $\phi_i$ onto some grid point $P$. The set of grid points $x_P$ and the resulting joint collocation matrix define a ``starting single-point'' grid, which we will seek to improve upon. Based on the interpretation of the doubles amplitudes as a four-electron quantity, we define an extension to this starting grid that accounts for the simultaneous contribution of pairs of grid points as,
\begin{align}\label{eq:joint-clloclation-matrix-pairs-of-gridpoints}
\begin{split}
\Delta Y^{PQ}_{ai} &\equiv (X_a^{P} + X_a^{Q})(X_i^{P} + X_i^{Q})  \\ 
&= Y_{ai}^{P} + Y_{ai}^{Q} + X_a^{P} X_i^{Q} + X_a^{Q} X_i^{P}
\end{split}
\end{align}
This ``pair point'' joint collocation matrix overlaps with the starting grid collocation matrix $\mathbf{Y}$, but now also includes products of collocation matrices $\mathbf{X}$ at unrelated grid points. This formalism does not include pair points in the most general way, since $X_a^{P} X_i^{Q}$ and $X_a^{Q} X_i^{P}$ are always paired in order to maintain overall separability. However, this separability property would be necessary for practical application in any case. We may then define a new joint collocation matrix, $\mathbf{Y^\prime}$, that is concatenation of the starting grid and pair point collocation matrices,
\begin{align}\label{eq:new_collocation}
    \mathbf{Y^\prime} = 
    \begin{bmatrix}
      \mathbf{Y} & \mathbf{\Delta Y}
    \end{bmatrix}.
\end{align}
with elements $Y^\prime_{(ai),(P,PQ,\ldots)}$. The task is to propagate the influence of $\mathbf{\Delta Y}$ through the rest of the LS-THC procedure. We must first define and invert the new metric matrix, which will be partitioned as
\begin{align}\label{eq:updated-S-matrix}
\mathbf{S^\prime} \equiv 
\begin{bmatrix}
      \mathbf{S} & \mathbf{Y}^{T}\mathbf{\Delta Y} \\
      \mathbf{\Delta Y} ^{T}\mathbf{ Y}  & \mathbf{\Delta Y} ^{T}\mathbf{  \Delta Y}
\end{bmatrix},
\end{align}
where in this notation,
\begin{align}\label{eq:metric}
    \mathbf{S} = \mathbf{Y}^T \mathbf{Y}
\end{align}

The new core matrix can then be determined by inserting the partitioned $Y^\prime$ and $S^\prime$ into the LS-THC fitting equations:
\begin{align}\label{eq:fitting}
    T_{PQ} &= \sum_{RS} \sum_{ijab} \left(S^{-1}\right)_{PR}
    Y^{R}_{ai}
    t^{ab}_{ij}
    Y^S_{bj}
    \left(S^{-1}\right)_{SQ} \nonumber \\
    \mathbf{T} &= \mathbf{S}^{-1} \mathbf{Y}^T \mathbf{t Y} \mathbf{S}^{-1}
\end{align}
with the doubles amplitudes stored as a matrix $\mathbf{t}$ with elements $t_{(ai),(bj)}$. Expanding the basis with pair points results in a modified core matrix,
\begin{align}\label{eq:T-with-pairs-of-grid-points}
    \mathbf{T^\prime} &= \mathbf{(S^\prime)}^{-1} (\mathbf{Y^\prime})^T \mathbf{ t Y^\prime} \mathbf{(S^\prime)}^{-1} \nonumber \\
    &=
\begin{bmatrix}
          \mathbf{S} & \mathbf{Y}^{T}\mathbf{\Delta Y} \\
      \mathbf{\Delta Y} ^{T} \mathbf{Y} & \mathbf{\Delta Y }^{T} \mathbf{ \Delta Y}
\end{bmatrix}
^{-1} 
\begin{bmatrix}
     \mathbf{Y} & \mathbf{\Delta Y}
\end{bmatrix}^{T}
\mathbf{t}
\begin{bmatrix}
     \mathbf{Y} & \mathbf{\Delta Y}
\end{bmatrix} 
\begin{bmatrix}
          \mathbf{S} & \mathbf{Y}^{T}\mathbf{\Delta Y} \\
      \mathbf{\Delta Y }^{T}\mathbf{Y}  & \mathbf{\Delta Y }^{T} \mathbf{ \Delta Y}
\end{bmatrix}
^{-1}
\end{align}
which, remarkably, is a completely closed-form equation thanks to the fact that LS-THC is entirely deterministic and does not involve any nonlinear fitting or iterative solutions of equations. Furthermore, the update $\mathbf{T}^\prime-\mathbf{T}$ can be efficiently obtained by using block inversion, especially when one pair grid point is added at a time.

Given this updated representation of $\hat{T}_2$, we may now begin to characterize how pairs of points contribute the the factorized correlation energy for a closed-shell system, split into ``Coulomb-like'' and ``exchange-like'' contributions,
\begin{align}\label{eq:ex-and-ec}
    \EC &= 2 \sum_{abij} g^{ij}_{ab}t^{ab}_{ij}  \\
    \EX &= -\sum_{abij}g^{ji}_{ab}t^{ab}_{ij}
\end{align} 
Inserting \eqref{eq:T-with-pairs-of-grid-points} into \eqref{eq:ex-and-ec} results in formulae for energies including both starting single grid points and pair grid points, 
\begin{align}
    \EC^\prime &= 2 \sum_{abij}\sum_{efmn}\sum_{PQRS} g^{ij}_{ab} (Y^\prime)^{P}_{ai} \left((S^\prime)^{-1}\right)^{PR} (Y^\prime)^{R}_{em} t^{ef}_{mn} (Y^\prime)^{S}_{fn}\left((S^\prime)^{-1}\right)^{SQ} (Y^\prime)^{Q}_{bj}  \nonumber \\
    &= 2\Tr\left[ \mathbf{g} \mathbf{Y^\prime} (\mathbf{S^\prime})^{-1}(\mathbf{Y^\prime})^T \mathbf{t} \mathbf{Y^\prime}(\mathbf{S^\prime})^{-1} (\mathbf{Y^\prime})^T \right] \label{eq:EC-LS-THC} \\
    \EX^\prime &= - \sum_{abij}\sum_{efmn}\sum_{PQRS} g^{ji}_{ab} (Y^\prime)^{P}_{ai} \left(S^{-1}\right)^{PR} (Y^\prime)^{R}_{em} t^{ef}_{mn} (Y^\prime)^{S}_{fn}\left(S^{-1}\right)^{SQ} (Y^\prime)^{Q}_{bj}, \nonumber \\
    &= -\Tr\left[ \mathbf{g}^\prime \mathbf{Y^\prime} (\mathbf{S^\prime})^{-1}(\mathbf{Y^\prime})^T \mathbf{t} \mathbf{Y^\prime}(\mathbf{S^\prime})^{-1} (\mathbf{Y^\prime})^T \right] \label{eq:EX-LS-THC}
\end{align} 
where the only difference between $\EC^\prime$ and $\EX^\prime$ is the ordering of the occupied orbitals on the ERIs, with $\mathbf{g}$ ordered as $g_{(ai),(bj)}$ and $\mathbf{g}^\prime$ ordered as $g^\prime_{(ai),(bj)} = g_{(aj),(bi)}$, respectively. Using block inversion based on the Woodbury matrix identity, we can arrive at a form which exposes the incremental energy contributed by one or more pair grid points,
\begin{align}
    \EC^\prime &= \left\{ 2\Tr \left[ \mathbf{g}\mathbf{Y} \mathbf{S}^{-1} \mathbf{Y}^{T}\mathbf{t} \mathbf{Y} \mathbf{S}^{-1} \mathbf{Y}^{T} \right] \right\} + \left\{ 4\Tr \left[ \mathbf{g} \mathbf{Y}\mathbf{S}^{-1}\mathbf{Y}^{T} \mathbf{t} \mathbf{C} \right] + 2\Tr \left[ \mathbf{g} \mathbf{C}^T \mathbf{t}\mathbf{C} \right] \right\} \nonumber \\
    &= \EC + \Delta \EC \label{eq:pair_energy} \\
    \mathbf{C} &= (\mathbf{Y}\mathbf{S}^{-1}\mathbf{Y}^{T} - \mathbf{I})\mathbf{\Delta Y} (\mathbf{\Delta Y} ^{T}  \mathbf{\Delta Y} - \mathbf{\Delta Y}^{T} \mathbf{Y} \mathbf{S}^{-1}\mathbf{Y}^{T}\mathbf{\Delta Y})^{-1} (\mathbf{\Delta Y})^{T} (\mathbf{Y}\mathbf{S}^{-1}\mathbf{Y}^{T} - \mathbf{I}) \label{eq:C}
\end{align}
where $\EC$ and $\Delta \EC$ refer to the LS-THC energy contributions from single grid points (the first term) and pair of grid points (the second two terms), respectively, and similarly for $\EX$. Equation \eqref{eq:pair_energy} (with the final working form given in \eqref{eq-factorized} in the Appendix) allows us to calculate the contributions from a large number of pairs of grid points at a time using level-3 BLAS operations, and thus provides an efficient means by which to determine how the LS-THC correlation energy responds to the inclusion of new pair points.

\subsection{Wavefunction Analysis via Weighted Subspace Score}

In this section we develop an independent methodology to determine the relative importance of a pair of grid points to the LS-THC representation of $\hat{T}_2$ via a weighted subspace score (WSS) which reflects the importance of pair points in a purely geometric sense. In WSS, the amplitude tensor, $t^{ab}_{ij}$ is cast into a square matrix $\mathbf{t}$ with entries $t_{(ai),(bj)} = t^{ab}_{ij}$, and then factorized using a singular value decomposition (SVD),
\begin{align}\label{eq:SVD}
    \mathbf{t} = \mathbf{U \Sigma} \mathbf{V}^T
\end{align}
where $\mathbf{\Sigma}$ is the diagonal singular value matrix and $\mathbf{U}$ ($\mathbf{V}$) is the matrix of left (right) singular vectors. Since $\mathbf{t}$ is symmetric, this may also be written as an eigenvalue decomposition (EVD), such that the eigenvalues are equal to the singular values and the eigenvectors are equal to the singular vectors, up to a sign which does not affect the weighted subspace analysis. Our goal is to determine how much of the original $\mathbf{t}$ matrix, expressed in the MO basis, is recovered within the LS-THC representation given a set of $n_g$ single grid points augmented with an additional set (typically only one) of pair grid points. The transformation into the rank-deficient LS-THC space is defined by the joint collocation matrix, $\mathbf{Y}$, defined in the previous section. The orthogonal basis of $\mathbf{Y}$ (or $\mathbf{\Delta Y}$) may be constructed via a QR decomposition:
\begin{align}
    \mathbf{Y} &= \mathbf{QR} \label{eq:Y_QR-decomp} \\
    \mathbf{\Delta Y} &= \mathbf{\Delta Q \Delta R}
\end{align}
where $\mathbf{Q}$ serves as the projector from the MO basis into the rank-deficient basis of single grid points, and $\mathbf{\Delta Q}$ serves the same role for pair grid points. As we are primarily interested in the contribution of only the pair points, captured in $\mathbf{\Delta Y}$, we measure the overlap of $\mathbf{\Delta Q}$ onto $\mathbf{t}$ after projecting out the single point space. The eigenvalues/singluar values provide a means of weighting the overlap, resulting in the formula,
\begin{align}
    \text{WSS} &= \frac{|\mathbf{U\Sigma V}(\mathbf{I}-\mathbf{QQ}^T)\mathbf{\Delta Q}|^2_F}{|\mathbf{\Sigma}|^2_F} \nonumber \\
    &= \frac{|\mathbf{\Sigma V}(\mathbf{I}-\mathbf{QQ}^T)\mathbf{\Delta Q}|^2_F}{|\mathbf{\Sigma}|^2_F}\label{eq:WSS-Y}
\end{align}
where the $F$ subscript indicates a Frobenius norm, $|\mathbf{A}|^2_F = \Tr[\mathbf{A}^T \mathbf{A}]$.
The WSS then provides a measure of how much any given pair point in $\mathbf{\Delta Y}$ contributes to the missing four-center components of the $\hat{T}_2$ kernel in \eqref{eq:t2-amplitude}.

\subsection{Computational Details}\label{sec:comp_details}

We preformed DF-MP2, -MP3, and -CCSD calculations on a set of ten molecules: adenine, thymine, the Watson--Crick adenine--thymine base-pair (AT-WC), a stacked adenine--thymine base pair (AT-ST), benzene, perfluorobutanoic acid (PFBA), a phenyl-substituted Criegee intermediate (Ph-Criegee, \ce{(C6H5)CHOO}), and linear alkanes $n$-hexane , $n$-octane, and $n$-decane. These calculations were performed with four basis sets---cc-pVDZ, aug-cc-pVDZ, cc-pVTZ(no f), and cc-pVTZ.\cite{Dunning-JCP-1989,Kendall-JCP-1992} The cc-pVTZ(no f) basis set excludes all $f$ functions from cc-pVTZ. All calculations were performed with RHF reference wavefunctions and with the frozen core approximation. Molecular geometries were optimized using B3LYP-D3/def2-TZVP\cite{Grimme-ChemR-2016,Weigend-PCCP-2005} in {\sc orca} version 5.0.3.\cite{Neese-JCP-2020} The correlated wavefunction calculations and LS-THC grid point generation were performed in a development version of {\sc cfour}\cite{Matthews-JCP-2020}. The parent grid for further analysis was taken as a standard SG-0 grid, pruned to a Cholesky tolerance of $\epsilon = 10^{-4}$. The LS-THC energy calculation, pair point energy partitioning, and pair point WSS analysis were implemented in a separate C++ program which is available in the Supplementary Information, along with more implementation details and additional scripts. For each molecule/basis set/method combination, we constructed starting grids with $\chi$ varying from 0.2 to 6.0 by truncating the grid basis obtained from the pivoted Cholesky decomposition. For each of these starting grids, we sampled 50,000 pairs of points from the parent grid (before truncation) with replacement, but excluding degenerate pairs (pairs with the same point repeated twice).

\section{Results}

\subsection{Single point grids}

\begin{figure}
    \centering
    \includegraphics[width = \textwidth]{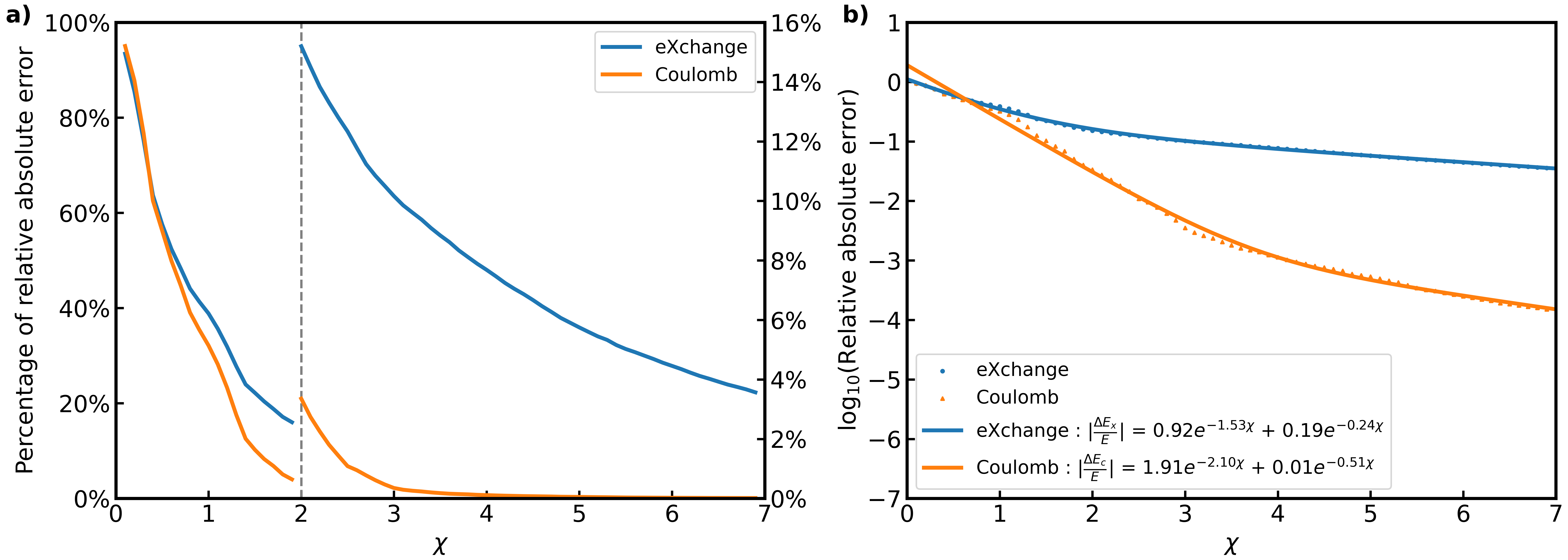}
    \caption{\textbf{a)} Percent relative absolute error of LS-THC $\EC$ ($\delta\EC$, orange) and $\EX$ ($\delta\EX$, blue) contributions to the CCSD/cc-pVTZ valence correlation energy of adenine. Note that the data to the left or right of the vertical dashed line corresponds to the $y$-axis on the same side. \textbf{b)} Bi-exponential fits of the of the relative absolute $\EC$ and $\EX$ errors with respect to $\chi$.}
    \label{fig:partial_energy_fitting}
\end{figure}

Previous work has demonstrated that the quality of Coulomb-like and exchange-like contributions to correlation energies differ substantially when the $\hat{T}_2$ amplitudes are THC decomposed.\cite{lee-JCTC-2020,Matthews-JCP-2021} We explore this phenomenon here across a wide range of molecules, basis sets, and wavefunctions as discussed in the previous section. In particular, we are interested in the rate of convergence of these correlation energy contributions as a function of the number of LS-THC grid points, chosen by the ordering given by the pivoted Cholesky procedure described in Ref.~\citenum{Matthews-JCTC-2020}. A typical plot of the convergence of the LS-THC errors in $\EC$ and $\EX$, as a fraction of the total canonical energy contributions, is given in Fig.~\ref{fig:partial_energy_fitting}. The relative errors (in percent), are defined as,
\begin{align}
    \delta \EC = \left| \frac{\EC(\chi)-\EC(\infty)}{\EC(\infty)} \right| \times 100\% \\
    \delta \EX = \left| \frac{\EX(\chi)-\EX(\infty)}{\EX(\infty)} \right| \times 100\%
\end{align}
where $\EC(\infty)$ and $\EX(\infty)$ are the canonical energies. To characterize this behavior, we deem the Coulomb-like correlation energy as ``converged'' when the LS-THC $\EC$ exhibits less than 0.1\% error from the canonical equivalent. This typically occurs for $\chi$ in the range from $\sim3.1$ to $\sim4.3$. We call the ratio $\chi$ at this point $\chi_C$. We can then measure the relative error in the exchange-like correlation energy at $\chi_C$ to determine the remaining errors arising from the LS-THC decomposition of $\hat{T}_2$.\footnote{While data is not presented here, note that $\EC$ and $\EX$ converge at similar rates when $\hat{T}_2$ is not approximated, i.e. the ``MP2a'' method of Ref.~\citenum{Matthews-JCP-2021}.}  

Further analysis is provided by fitting the error of $\EC$ and $\EX$ as a function of $\chi$. Errors of both $\EC$ and $\EX$ display a bi-exponential dependence on the ratio of grid points to active orbitals, with a relatively clear transition from ``fast'' convergence, where the energy error decays rapidly with each additional grid point, to ``slow'' convergence, where each added grid point accounts for less and less of the remaining error. To some extent this should be taken as an indication that the selection of the grid points via Cholesky pruning does a good job of ordering the grid points relative to their importance in the wavefunction, despite the fact that no explicit considerations of the energy or correlated wavefunction are made in that procedure. Here we observe an even more significant difference between the behavior of $\EC$ and $\EX$. The $\EC$ LS-THC contributions to the correlation energy generally converge (to within $0.1\%$ of the canonical values) before or at the same $\chi$ for which the decay transitions from the fast to the slow convergence regimes. In contrast, the $\EX$ LS-THC contributions have an average error of approximately 4.89\%, 6.64\%, and 7.46\% at $\XC$ for the MP2, MP3, and CCSD methods, respectively.

\begin{figure}
    \centering
    \includegraphics[width = 0.8\textwidth]{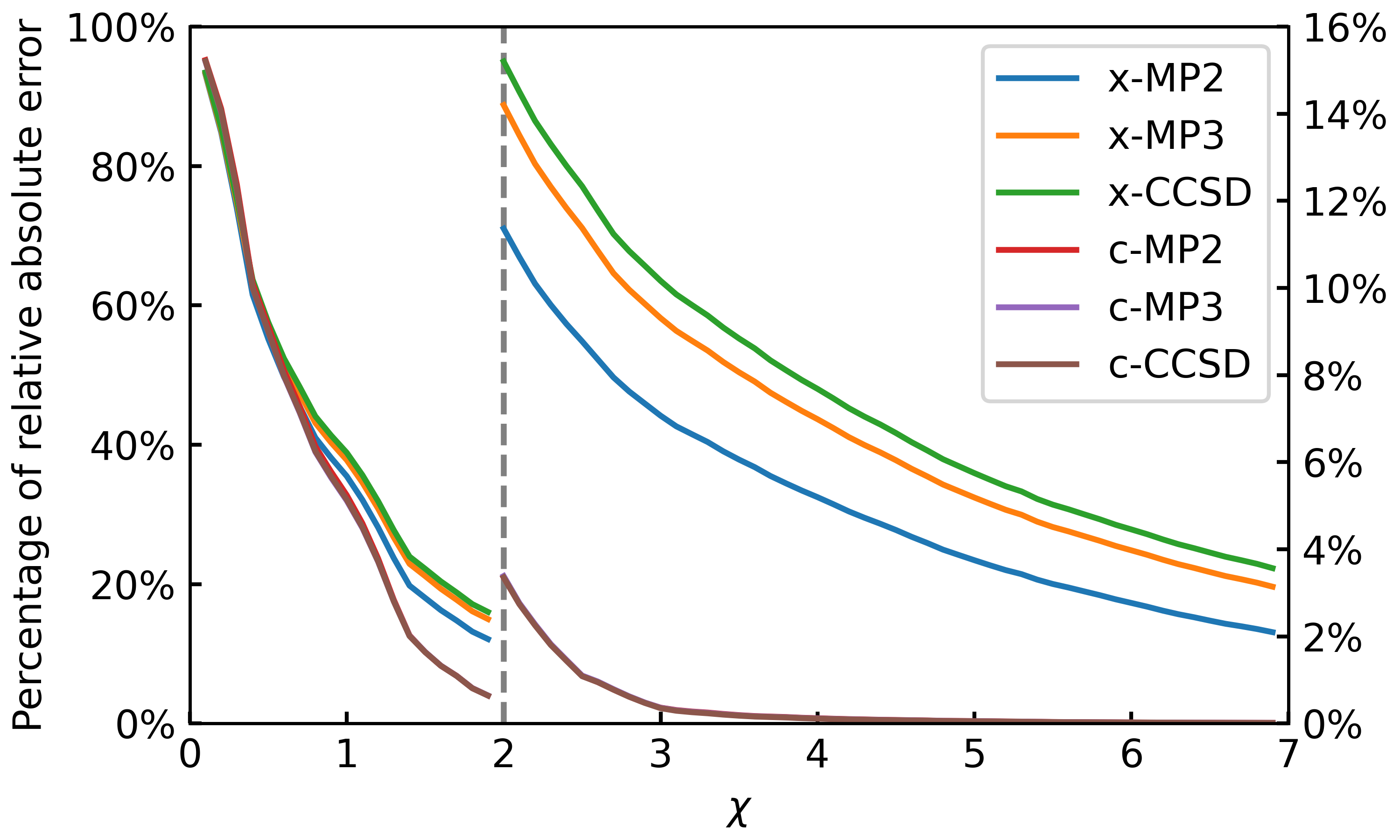}
    \caption{Percent error of LS-THC $\EC$ ($\delta\EC$: red, purple, and brown) and $\EX$ ($\delta\EX$: blue, orange, and green) for adenine calculated with MP2, MP3, and CCSD using the cc-pVTZ basis set.}
    \label{fig:partial_energy_methods}
\end{figure}

This behavior is very consistent across all the molecules studied here. There are minor differences depending upon the basis set and wavefunction in question, which are worth brief discussion. For instance, Fig. \ref{fig:partial_energy_methods} displays the percentage of relative error in $\EC$ and $\EX$ for adenine calculated with MP2, MP3, and CCSD with the cc-pVTZ basis set. While the profile of $\EC$ as a function of $\chi$ is almost identical for all three wavefunctions, $\EX$ requires monotonically increasing numbers of grid points with more correlated wavefunctions (MP2 $<$ MP3 $<$ CCSD). This trend holds for all molecules, see Table~\ref{table:partial_energy_methods}, and over all basis sets studied here, see SI. This trend may be partially explained by the eigenvalue structure of the $\hat{T}_2$ amplitudes used in each method. For example, as noted in Ref.~\citenum{Parrish-JCP-2019}, the first-order doubles amplitudes used in MP2 are strictly negative definite, while for MP3, and to a slightly greater extent, CCSD, the doubles amplitudes pick up a positive branch in their spectrum.

\begin{table}
\begin{threeparttable}
\centering
\caption[short]{Values of $\chi_C$ and relative absolute error $\delta\EX$ (in \%) at the $\EC$ convergence point for the MP2, MP3, and CCSD/cc-pVTZ methods.} \label{table:partial_energy_methods}
\begin{tabular}{p{2cm} *{6}{p{1.5cm}<{\centering}}}
\midrule 
 & \multicolumn{2}{c}{MP2}                    & \multicolumn{2}{c}{MP3}                    & \multicolumn{2}{c}{CCSD} \\ 
\cmidrule(r){2-7}

 &  $\chi_C$  & $\delta\EX$ & $\chi_C$ &  $\delta\EX$ &  $\chi_C$  &  $\delta\EX$  \\
\midrule 
Adenine       & 4.30       & \emph{4.72}       & 4.20         & \emph{6.58}\       & 4.20         & \emph{7.24}         \\
AT-ST         & 4.00       & \emph{5.25}       & 4.00         & \emph{7.10}\       & 3.90         & \emph{8.07}         \\
AT-WC         & 4.10       & \emph{4.90}       & 4.10         & \emph{6.64}\       & 4.00         & \emph{7.54}         \\
Benzene       & 3.40       & \emph{5.51}       & 3.30         & \emph{7.48}\       & 3.30         & \emph{8.15}         \\
PFBA          & 3.20       & \emph{4.23}       & 3.20         & \emph{6.02}\       & 3.20         & \emph{6.86}         \\
Ph-criegee    & 3.50       & \emph{5.44}       & 3.50         & \emph{7.25}\       & 3.40         & \emph{8.28}         \\
Thymine       & 3.60       & \emph{4.97}       & 3.60         & \emph{6.77}\       & 3.60         & \emph{7.59}         \\
Hexane        & 3.20       & \emph{4.58}       & 3.10         & \emph{6.23}\       & 3.10         & \emph{6.93}         \\
Octane        & 3.30       & \emph{4.61}       & 3.30         & \emph{6.10}\       & 3.20         & \emph{6.99}         \\
Decane        & 3.30       & \emph{4.70}       & 3.30         & \emph{6.23}\       & 3.30         & \emph{6.92}         \\
\midrule
Mean          & 3.59       & \emph{4.89}       & 3.56         & \emph{6.64}\       & 3.52         & \emph{7.46}         \\
Std. Dev.     & 0.40       & \emph{0.41}       & 0.40         & \emph{0.51}\       & 0.39         & \emph{0.55}          \\                  
\midrule
\end{tabular} 
\end{threeparttable}
\end{table}

\begin{table}
\begin{threeparttable}
\centering
\caption[short]{Values of $\chi_C$ and relative absolute error $\delta\EX$ (in \%) for CCSD wavefunctions at the $\EC$ convergence point: cc-pVDZ, aug-cc-pVDZ, cc-pVTZ(no~f), and cc-pVTZ basis sets.} \label{table:partial_energy_basis_set}
\begin{tabular}{ p{2cm} *{8}{p{1.2cm}<{\centering}}}
\midrule 
 & \multicolumn{2}{c}{cc-pVDZ}  & \multicolumn{2}{c}{aug-cc-pVDZ}  & \multicolumn{2}{c}{cc-pVTZ(no~f)}  & \multicolumn{2}{c}{cc-pVTZ}\\ 
\cmidrule(r){2-9}

 &  $\chi_C$  & $\delta\EX$  & $\chi_C$ &  $\delta\EX$  &  $\chi_C$  & $\delta\EX$  &  $\chi_C$  &  $\delta\EX$ \\
\midrule 
Adenine    & 4.20 & \emph{5.22} & 4.30 & \emph{4.76} & 4.10 & \emph{6.51} & 4.20 & \emph{7.24} \\
AT-ST      & 4.40 & \emph{5.02} & 4.10 & \emph{5.42} & 4.00 & \emph{6.81} & 3.90 & \emph{8.07} \\
AT-WC      & 4.20 & \emph{5.13} & 4.10 & \emph{5.26} & 4.00 & \emph{6.51} & 4.00 & \emph{7.54} \\
Benzene    & 3.80 & \emph{5.17} & 3.20 & \emph{5.75} & 3.50 & \emph{6.41} & 3.30 & \emph{8.15} \\
PFBA       & 3.80 & \emph{4.00} & 3.20 & \emph{5.01} & 3.40 & \emph{5.59} & 3.20 & \emph{6.86} \\
Ph-Criegee & 4.00 & \emph{5.07} & 3.40 & \emph{5.81} & 3.70 & \emph{6.54} & 3.40 & \emph{8.28} \\
Thymine    & 4.00 & \emph{4.80} & 3.50 & \emph{5.54} & 3.70 & \emph{6.38} & 3.60 & \emph{7.59} \\
Hexane     & 3.70 & \emph{4.00} & 3.00 & \emph{4.77} & 3.20 & \emph{5.32} & 3.10 & \emph{6.93} \\
Octane     & 3.80 & \emph{4.11} & 3.10 & \emph{4.80} & 3.20 & \emph{5.48} & 3.20 & \emph{6.99} \\
Decane     & 3.80 & \emph{4.19} & 3.10 & \emph{4.92} & 3.20 & \emph{5.61} & 3.30 & \emph{6.92} \\
\midrule
Mean       & 3.97 & \emph{4.67} & 3.50 & \emph{5.20} & 3.60 & \emph{6.12} & 3.52 & \emph{7.46} \\
Std. Dev.  & 0.23 & \emph{0.53} & 0.49 & \emph{0.41} & 0.35 & \emph{0.55} & 0.39 & \emph{0.55} \\
\midrule
\end{tabular} 
\end{threeparttable}
\end{table}

Basis set dependence follows a similar pattern, see Table~\ref{table:partial_energy_basis_set}, with larger basis sets presenting more relative error in $\EX$ when $\EC$ has reached convergence. As with wavefunction dependence, the ratio of grid points to active orbitals required to converge $\EC$ is very stable across all basis sets save cc-pVDZ: the average $\chi_C$ for aug-cc-pVDZ, cc-pVTZ(no~f), and cc-pVTZ have only minimal differences and are well within one standard deviation of one another. It is perhaps not surprising that it requires a slightly larger ratio of grid points to active orbitals to converge $\EC$ for a cc-pVDZ basis set. It is conceivable that there is simply a base number of spatial grid points required to represent the wavefuction (akin to a minimally spanning orbital basis set), and cc-pVDZ is such a small basis set that this inflates $\chi_C$ even if the number of grid points actually used in the calculation is relatively small.  

In all regards, we observe that LS-THC decomposition of $\EX$ is manifestly dissimilar from $\EC$. Further we note that this problem appears to be specific to the LS-THC compression: other techniques such as rank-reduced CCSD\cite{Parrish-JCP-2019}, nonlinear THC\cite{Schutski-JCP-2017,Gruneis-JCP-2017,Hohenstein-JCP-2022}, and LS-THC where only ERIs are approximated (e.g. MP2a/MP3a\cite{Matthews-JCP-2021} or LS-THC-(EOM-)CC2\cite{Hohenstein-JPCB-2013} do not exhibit such differences between exchange-like and Coulomb-like interactions. Understanding and controlling this behavior is critical to the success of LS-THC methods for complex wavefunctions such as MP3 and CCSD. Motivated by the real-space interpretation of $\hat{T}_2$ discussed in the Introduction, we now turn to an analysis of the spatial relationships between pairs of grid points in an attempt to tease out more information about the relationship between LS-THC errors and the grid points that makeup the representation of these wavefunctions.

%
%
%
%

\subsection{Spatial Analysis of Pair Grid Points}

In this section we attempt to probe the nature of the slow convergence of the exchange energy in LS-THC. From the perspective of the real space doubles amplitude kernel, LS-THC is missing some portion of pair point dependence in the wavefunction, which can be understood through a quadrature argument,
\begin{align}
    t_{ij}^{ab} &= \int \phi_a(r_1) \phi_i(r_1^{'}) \hat{t}(r_1, r_2, r_1^{'}, r_2^{'})  \phi_b(r_2) \phi_j(r_2^{'}) dR \nonumber \\
    &\approx \sum_{PQRS} X_a^{P} X_i^{Q} \hat{t}(x_P,x_R,x_{Q},x_{S})X_b^{R} X_j^{S} \label{eq:t2-quadrature}
\end{align}
assuming the quadrature weights are absorbed into $\mathbf{X}$. Clearly, the standard LS-THC approximation neglects contributions with $P\ne Q$ or $R\ne S$, and a complete description should allow for distinct pairs of points describing the electronic distributions, though what form this dependence should take is unknown. In the previous section we introduced one such ansatz which adds a joint collocation matrix with entries $\Delta Y_{ai}^{PQ}$, and thus provides a degree of pair point dependence. In this section, we examine the relationships between the spatial distribution of pairs of points and their estimated ``importance'', calculated using the $\Delta \EC$/$\Delta \EX$ energy analysis and WSS values derived above. From the 50,000 pair points sampled for each method/molecule/basis set/$\chi$ combination we construct a histogram of average pair point energy/WSS binned by the distance between the points in each pair.

\begin{figure}[t]
    \centering
    \includegraphics[width=\textwidth]{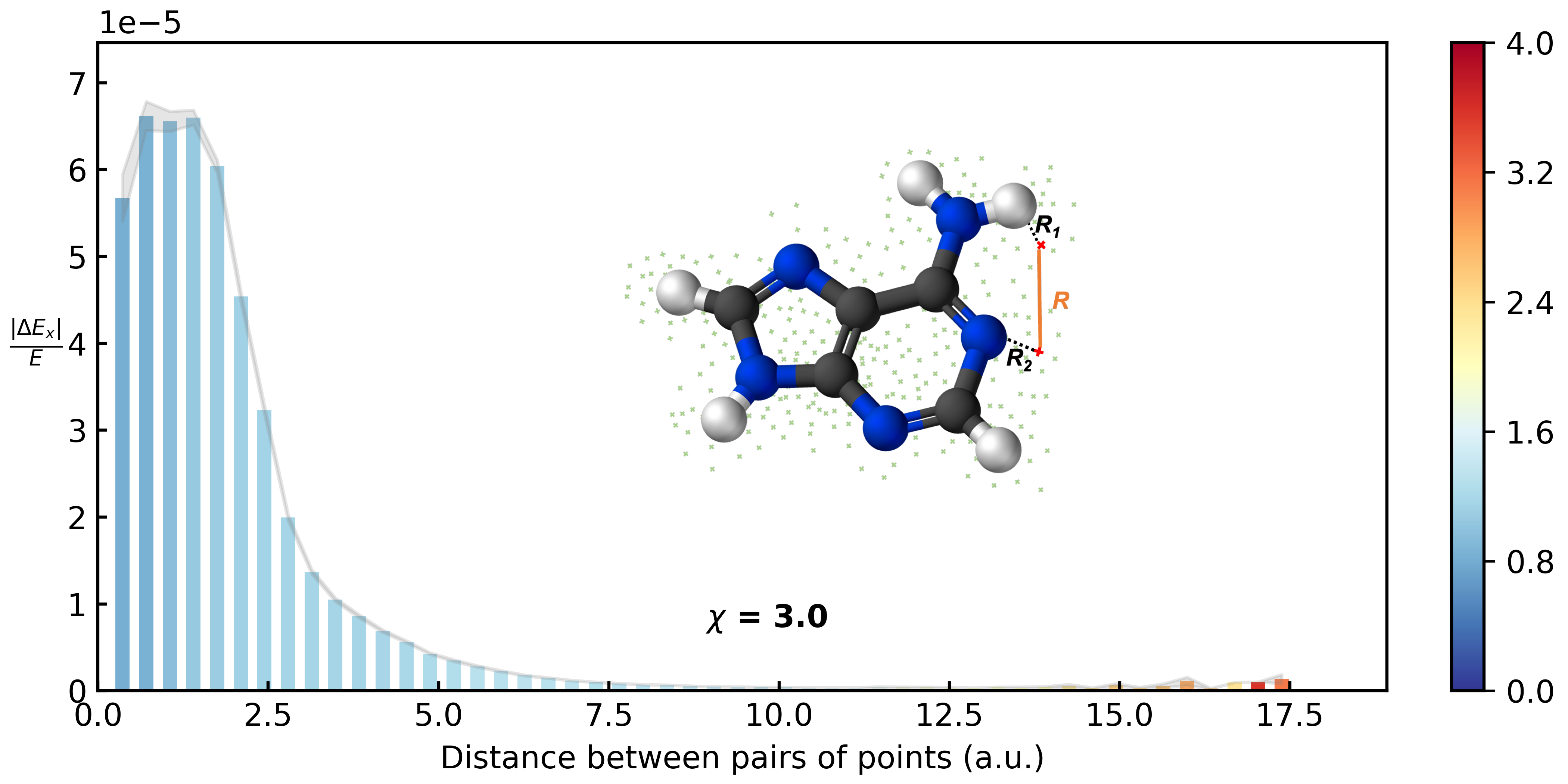}
    \caption{A typical ``spatial histogram'' plot: the $y$-axis represents average relative contributions $|\Delta\EX|/E$ (with $E$ being the total canonical correlation energy) for pair points added to adenine CCSD/cc-pVTZ with $\chi=3.0$. The $x$-axis is binned by inter-pair distance $R$ and bins are color-coded by the average distance of the pair points to their nearest atoms [$(R_1+R_2)/2$]. The gray shaded region indicates the standard deviation estimated by bootstrap sampling.}
    \label{fig:pairs_of_grid_points_example}
\end{figure}

We will illustrate this analysis in the context of a concrete example, as these plots are the primary means by which the main findings of this manuscript may be understood. Consider Fig.~\ref{fig:pairs_of_grid_points_example}, which is obtained from an LS-THC-CCSD/cc-pVTZ calculation of adenine, with $\chi = 3.0$, which provides a single-point starting grid of 1080 points. Starting with this single-point grid, we wish to determine how different pairs of points contribute to the description of the correlation energy/wavefunction that is missing from the single point space alone. Take, for example, the two points indicated by red x's in Fig. \ref{fig:pairs_of_grid_points_example}. Let us call these points $x_1$ and $x_2$, where point $x_1$ is of distance $R_1$ to its closest atom, point $x_2$ is of distance $R_2$ from its closest atom (which may be different), and the points are distance $R$ from one another. Then, the contribution of this $x_1,x_2$ pair to the energy/wavefunction that is not already accounted for in by the single point grid is determined by either the energy partitioning or WSS analysis described above, the result of which is recorded as a sample in the histogram bin covering the pair distance $R$. Results are averaged within each bin, with the energy or WSS scores represented as the bin height, and the mean closest atom distances $(R_1+R_2)/2$ as the bin color, with a gray shaded region indicating the standard deviation determined via bootstrap sampling with 5000 replicates.\cite{Efron-SS-1986} The end result of this is Fig.~\ref{fig:pairs_of_grid_points_example}, where the y-axis indicates roughly how important pairs of points a particular distance away from one-another are for the LS-THC description of the energy/wavefunction, and the color coding indicates roughly how close the pairs of points in a particular bin are to nearby atoms. We denote this type of plot a ``spatial histogram''. Fig.~\ref{fig:pairs_of_grid_points_example} already illustrates the typical pattern we observe across a wide range of molecules, methods, and basis sets: once single point grid saturation is reached (around $\chi=\chi_C$), the remaining important contributions to $\EX$ come form pair points which are close together ($R \lesssim 4$ a.u.) and are close to atoms ($R_1,R_2 \sim 1.5$ a.u.). 

\begin{figure}
    \centering
    \includegraphics[width = \textwidth]{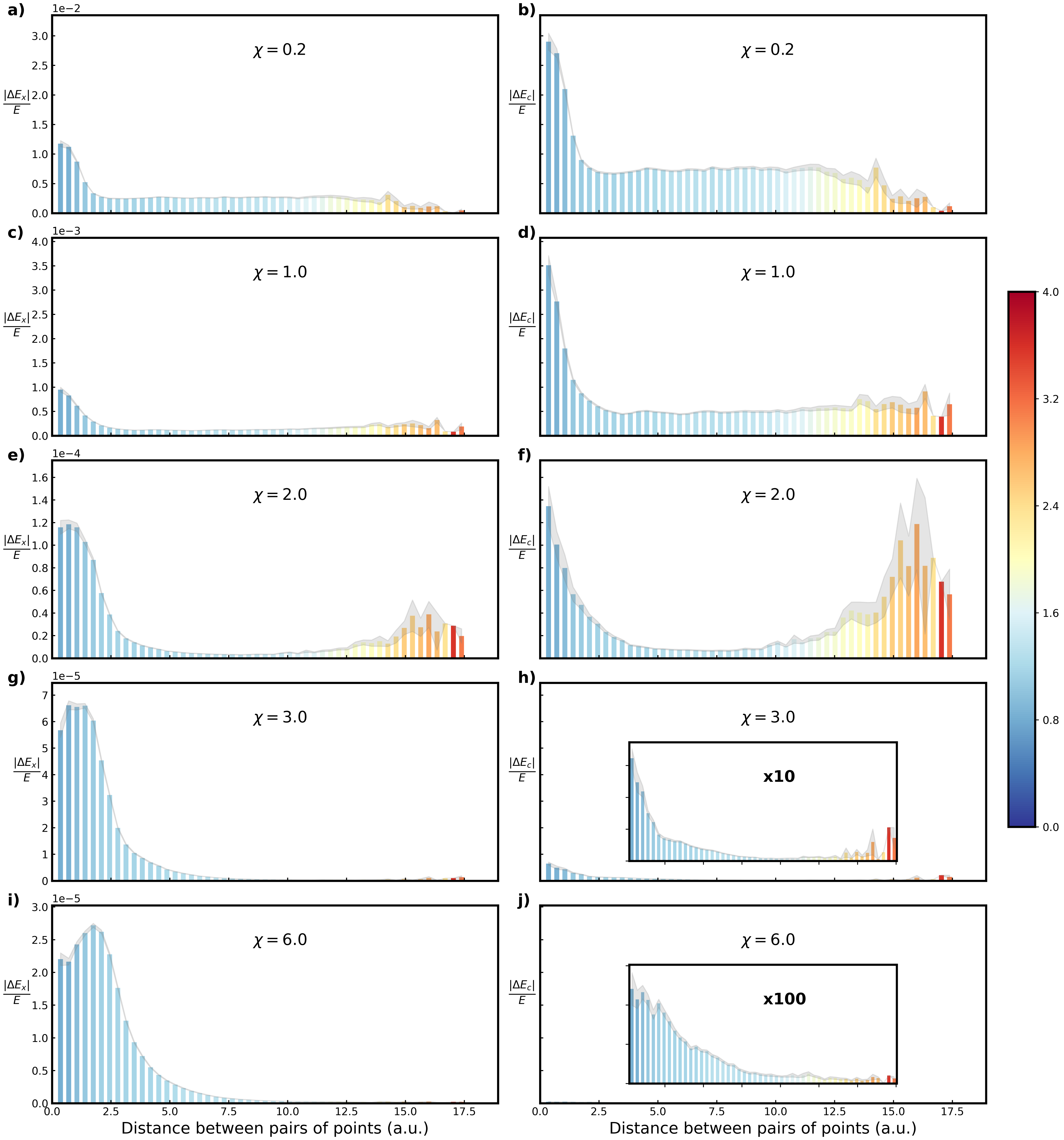}
    \caption{Spatial histograms of $\Delta\EX/E$ (left column) and $\Delta\EC/E$ (right column) for adenine CCSD/cc-pVTZ, where $E$ is the total canonical correlation energy. See Fig.~\ref{fig:pairs_of_grid_points_example} for details. $\chi$ increases from top to bottom, progressively saturating the singe point space. The inset plots in \textbf{h} and \textbf{j} are enlarged 10x and 100x, respectively.}
    \label{fig:pairs_of_grid_point_with_various_chi_X_and_C}
\end{figure}

The evolution of this phenomenon with the size of the starting single point grid is illustrated in Fig.~\ref{fig:pairs_of_grid_point_with_various_chi_X_and_C}, along with a comparison of the exchange-like ($\Delta\EX$) and Coulomb-like ($\Delta\EC$) contributions at each grid size. At small values of $\chi$ (0.2--1.0, Fig.~\ref{fig:pairs_of_grid_point_with_various_chi_X_and_C}a--d), the spatial dependence of pair points with the largest energy contributions to $\EC$ and $\EX$ have a similar spatial distribution: pairs of points less than 2.5 a.u. apart are disproportionately more important, with a fairly even spread of contributions for points that sit between 2.5 and 12.5 a.u. apart. Further, there is a distinct trend that the pair points with small $R$ (on the left of each plot) also tend to have smaller average distances to their nearest atoms (shaded darker blue). As the distance between the points grows (towards the right side of each plot) the average distance of these points to the nearest atom grows as well (shaded yellow to red), as does the uncertainty in the computed average (indicated by the width of the gray shaded region). These grid points in the diffuse region have an out-sized contribution to $\EC$ and $\EX$ due to the large grid weights (roughly proportional to the ``volume'' spanned by each point), but the extremely small number of such points makes them irrelevant to the total energy. The large grid weights also cause such points to be chosen during the Cholesky pruning procedure for inclusion in the starting grid (in Fig.~\ref{fig:pairs_of_grid_point_with_various_chi_X_and_C} around $\chi = 3$) after which the pair point contribution goes to essentially zero. Across our calculations we observe an average $|\EC(\infty)|:|\EX(\infty)|$ ratio of $\sim 3:1$. Accounting for the difference in magnitude of the two components further highlights that the energy contributions of pair points in the small grid regime ($\chi < 2$) are almost identical.

\begin{figure}[t]
    \centering
    \includegraphics{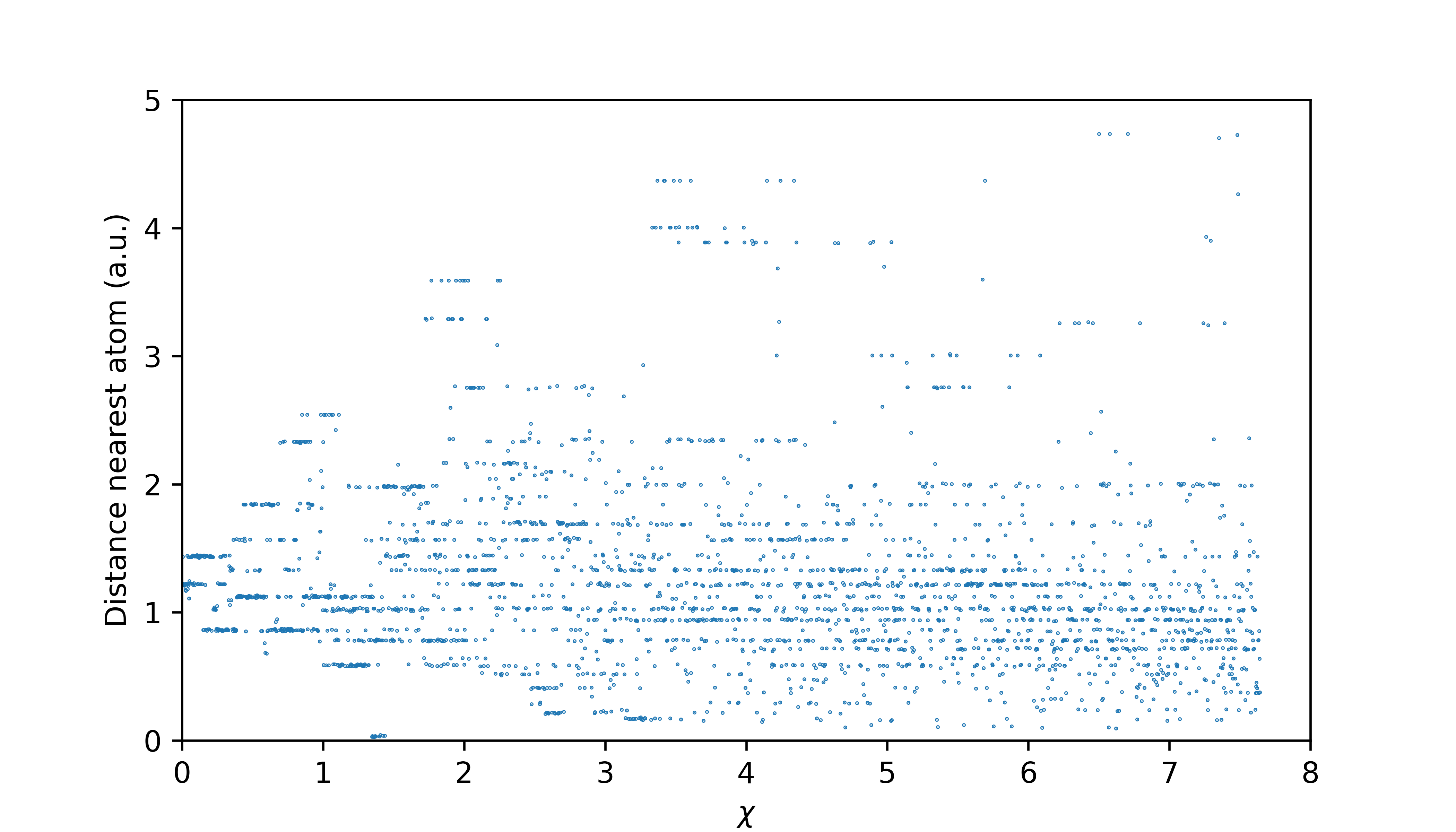}
    \caption{Distance of grid points to the nearest atom (a.u.), as points are selected by the pivoted Cholesky decomposition procedure\cite{Matthews-JCTC-2020} with increasing values of $\chi$. The points plotted at a particular value of $\chi$ indicated newly chosen points, and not the cumulative grid points so far selected. The grid and LS-THC metric matrix were calculated for adenine with a cc-pVTZ basis. }
    \label{fig:distance_between_grid_to_nearest_atom}
\end{figure}

However, as the number of starting grid points increases between $\chi=1$ and $\chi=3$, qualitative differences between the spatial dependence of $\EC$ and $\EX$ begin to emerge. Importantly, these values of $\chi$ correspond to the region where the single point grid alone nears convergence in $\EC$ (within 0.35\% of the final $\EC$ values by $\chi=3$), but $\EX$ is still missing more than 10\% of its final value, see Fig.~\ref{fig:partial_energy_fitting}a. This is also the region of transition between ``fast'' and ``slow'' convergence of $\EX$ in the LS-THC single point basis, see Fig.~\ref{fig:partial_energy_fitting}b. At $\chi=2$, both $\EC$ and $\EX$ show a pronounced decrease in the importance of pair-points at ``medium''  distances from one another ($3.0 < R < 12.5$ a.u.), leading to a ``bathtub''-shaped curve. This effect is mostly due to the order in which points are selected for inclusion in the single point grid (Fig.~\ref{fig:distance_between_grid_to_nearest_atom}) and the large number of points in this region. As noted above, the contributions of long-range pairs of points ($ R \gtrsim 12.5$ a.u.) mostly disappears by $\chi = 3$. However, the dependence of $\EX$ on short-range pairs of points condenses into a peak centered between 1 and 3 a.u., and this feature decays very slowly as the single point space is increased, despite the fact that this region already contains a dense concentration of single grid points (see Fig.~\ref{fig:distance_between_grid_to_nearest_atom}). In fact, by $\chi=3$ the vast majority of the LS-THC error for the CCSD/cc-pVTZ valance correlation energy of adenine is accounted for by only 15\% ($R<3$ a.u.) of the 50,000 pairs of points examined. 

\begin{figure}
    \centering
    \includegraphics[width = \textwidth]{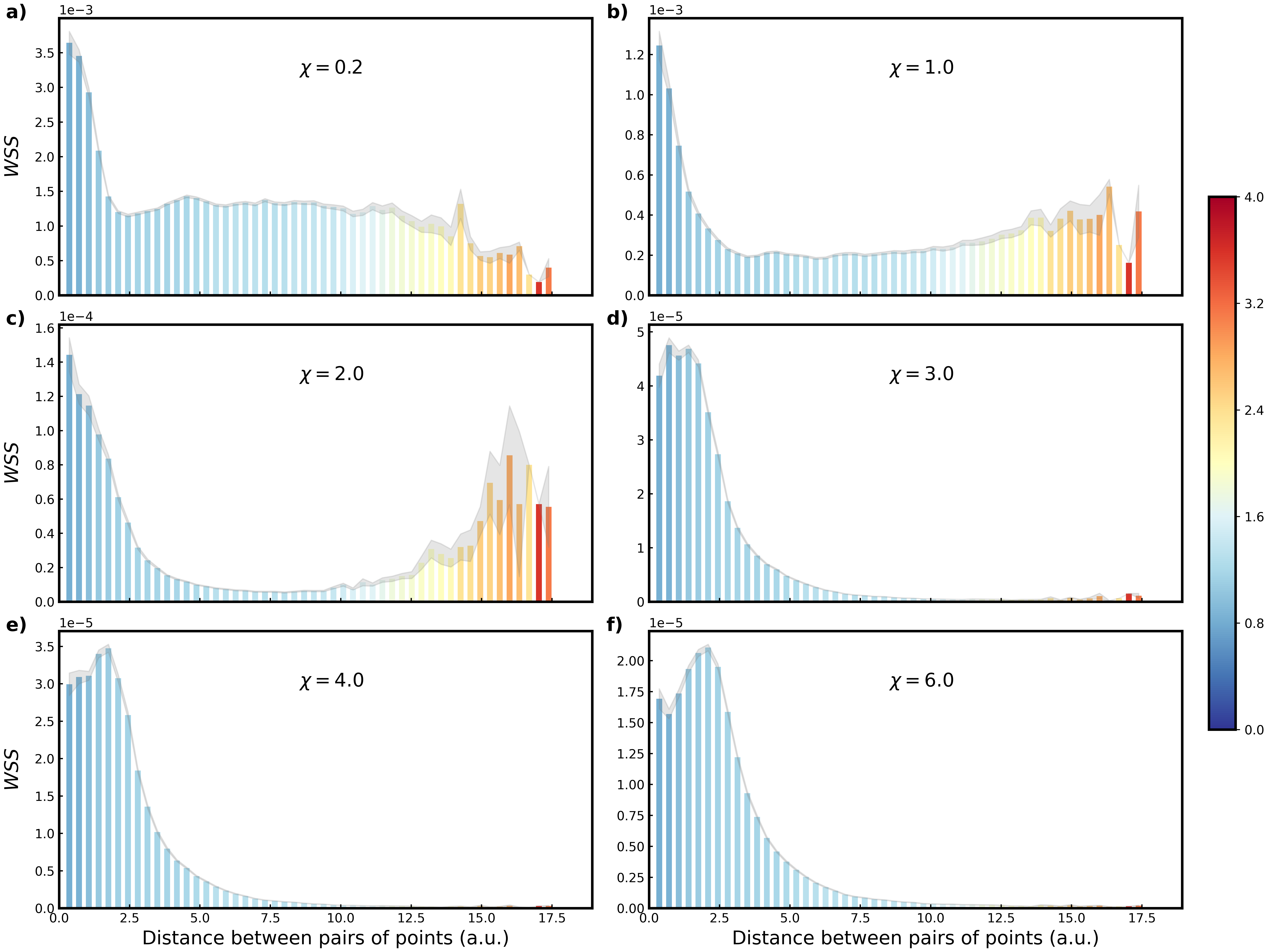}
    \caption{Spatial histogram of weighted subspace scores (WSS) of pair points with respect to $\hat{T}_2$ amplitudes calculated for adenine with CCSD/cc-pVTZ. See Fig.~\ref{fig:pairs_of_grid_points_example} for details.}
    \label{fig:pairs_of_grid_points_WSS_with_various_chi}
\end{figure}


\begin{table}
\centering
\caption[short]{Center and width of a student's $t$-distribution fit to exchange energy analysis and WSS analysis accumulated across all molecules, methods, and basis sets (excluding cc-pVDZ) as a function of the ratio of LS-THC grid points to the total number of basis functions ($\chi$). Values (in a.u.) are reported as the average followed by the standard deviation, with the source of the correlation feature indicated in parenthesis.} \label{table:peak_postion_of_Fval_and_exchange}
\begin{tabular}{*{5}{p{2.9cm}<{\centering}}}
\toprule
 $\chi$ & $R_0(\EX)$ & $R_0(\text{WSS})$ &  $\sigma(\EX)$ &  $\sigma(\text{WSS})$\\
\midrule 
3.0 & 1.22 $\pm$ 0.12 & 1.18 $\pm$ 0.17 & 1.39 $\pm$ 0.20 & 1.56 $\pm$ 0.21 \\
4.0 & 1.40 $\pm$ 0.17 & 1.47 $\pm$ 0.15 & 1.47 $\pm$ 0.22 & 1.55 $\pm$ 0.24 \\
5.0 & 1.56 $\pm$ 0.19 & 1.67 $\pm$ 0.11 & 1.54 $\pm$ 0.20 & 1.61 $\pm$ 0.24 \\
6.0 & 1.64 $\pm$ 0.23 & 1.78 $\pm$ 0.11 & 1.62 $\pm$ 0.23 & 1.72 $\pm$ 0.29 \\
\bottomrule
\end{tabular} 
\end{table}

Importantly, an identical spatial signature arises in the pair-point dependence of the $\hat{T}_2$ weighted subspace scores, see Fig.~\ref{fig:pairs_of_grid_points_WSS_with_various_chi}. As we will discuss in the following sections, this ``correlation feature'' displays only slight variations across basis sets, ground state correlated wavefunctions, and molecular systems. To obtain some quantitative comparisons, these correlation features were fit to a Student $t$-distribution defined as
\begin{align}\label{eq:$t$-distribution}
    P(R; R_0, \sigma, \nu, a ) = 
    \frac{a}{\sigma\sqrt{\pi \nu}}
    \frac{\Gamma\left(\frac{1}{2}(\nu + 1)\right)}
    {\Gamma\left(\frac{1}{2}\nu\right)}
    \left(1 + \frac{(R - R_0)^2}{\sigma^2\nu} \right)^{-\frac{1}{2}\left(\nu + 1 \right)},
\end{align}
where $\Gamma$ is the Gamma function, $R$ is the inter-pair distance, $R_0$ is the peak-center, $\nu$ is the number of degrees of freedom (taken as a fitting parameter here), $a$ is a constant, and $\sigma$ is a scale factor corresponding to the width of the distribution. The $t$-distribution is identical to the normal distribution when $\nu \rightarrow \infty$, where $\sigma$ takes on its usual meaning; in our fits we find $\nu$ typically in the range of 2--5. The use of the $t$-distribution for fitting should not be taken as an assumption about how the shape of the feature arises from the underlying physics. While we obtained good fits using this function, it is essentially an arbitrary broadening function. Across all species, wavefunctions, and basis sets, we found that this fit resulted in $R^2$ no worse than 0.98, and generally was a good fit for all the correlation features to which it was applied.

The parameters we are most interested in are $R_0$ and $\sigma$, the peak location and (roughly) peak width predicted by the fit. Table~\ref{table:peak_postion_of_Fval_and_exchange} reports these values averaged across all molecules, basis sets (save cc-pVDZ, for reasons discussed later), and wavefunctions as a function of $\chi$. A full tabulation of these fits may be found in the Supporting Information, including covariance matrices of the resulting parameters. Notably, $R_0$ has a very small standard deviation, and the peak position shifts to longer inter-pair distances as the coverage of the single point space increases (larger $\chi$). This shift is likely a consequence of the relative ordering of the single points via the Cholesky fitting, as grid points very close to the atoms are selected at larger values of $\chi$, see Fig.~\ref{fig:distance_between_grid_to_nearest_atom}. Likewise, $\sigma$ (related to the width of the features) has very small deviations across species, basis, and wavefunction. As with $R_0$, $\sigma$ grows slightly larger with increasing $\chi$, which likely results from the shift in $R_0$ to longer inter-pair distances causing less compression of the peak at small $\chi$.  

As stated above, one of the most salient properties of this correlation feature is its consistency across species, wavefunctions, and basis, as reflected in the small standard deviation of $R_0$ and $\sigma$ in the accumulated fits of both $\EX$ and WSS, on top of the similarity of the correlation feature between those two quantities. We explore the limited dependence of the correlation feature on these factors in the following subsections, and provide some discussion of this phenomenon.


\subsubsection{Trends Across Molecules}\label{sub:sub:pair_molecules}

\begin{figure}[t]
     \centering
     \includegraphics[width = \textwidth]{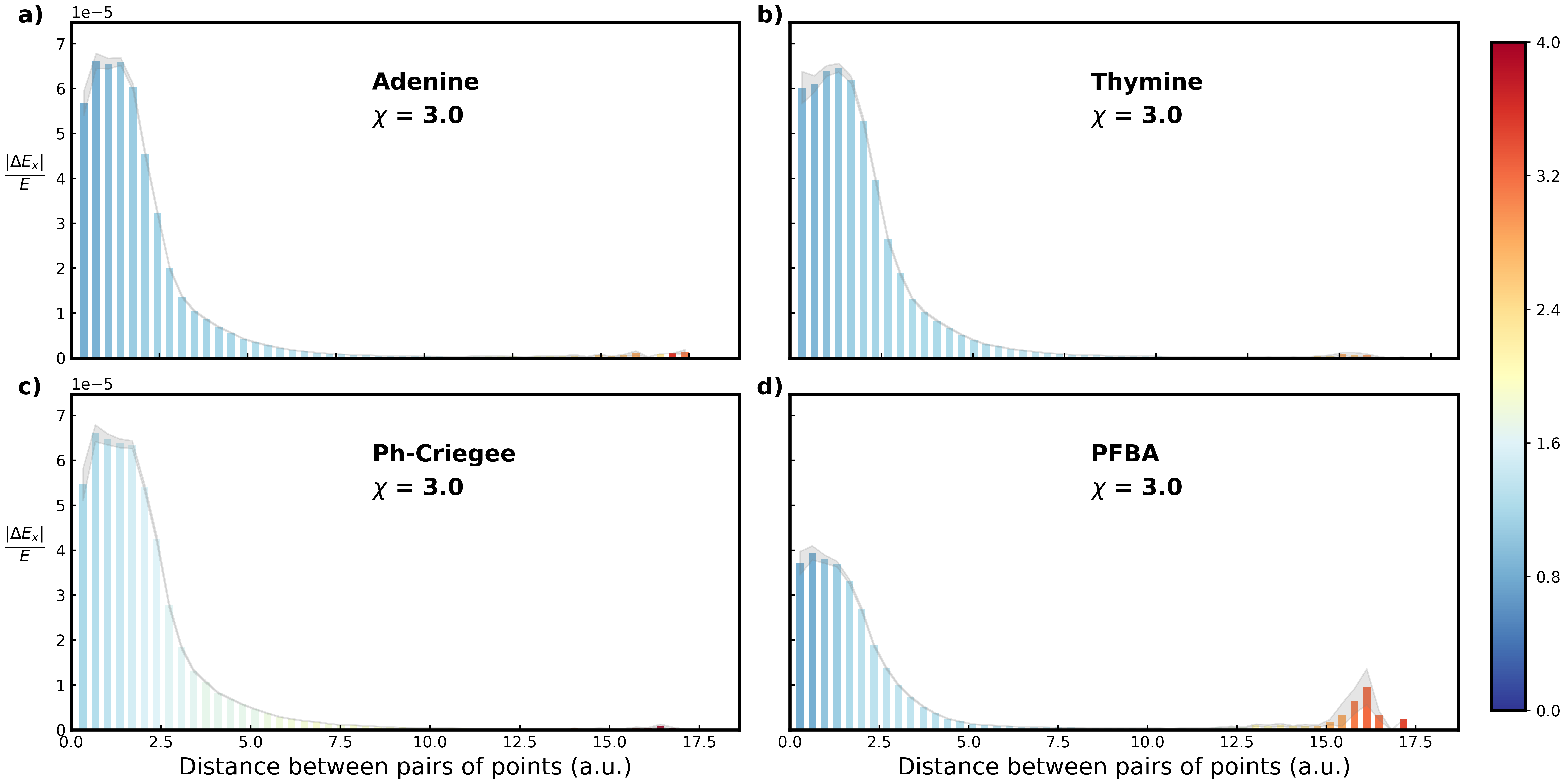}
     \caption{Spatial histograms of $|\Delta\EX|/E$ for various molecular systems at the CCSD/cc-pVTZ level. See Fig.~\ref{fig:pairs_of_grid_points_example} for details.}
     \label{fig:Pairs_of_grid_point_with_various_molecule}
 \end{figure}

A striking quality of this correlation feature is its consistency across the set of ten molecules studied here. Fig.~\ref{fig:Pairs_of_grid_point_with_various_molecule} displays the dependence of the magnitude of pair points contributions to $\EX$ to the valance CCSD/cc-pVTZ correlation energy for adenine, thymine, Ph-Crigee, and PFBA. The general shape and peak position of the correlation feature of these four species is largely identical, despite the wide array of bonding motifs, functional groups, dispersion interactions, and multireference character of these molecules. The stability of the spatial feature is tabulated further in Table~\ref{table:pairs_of_grid_points_dependency_to_molecules_R0_and_R0_width}, which reports the features of the $t$-distribution fits for all ten species, accumulated over the various wavefunctions and basis sets (again, excluding cc-pVDZ) studied here. The standard deviation of both the $R_0$ and $\sigma$ of these fits is remarkably small, further illustrating how consistent these features are between molecules.

There are some minor variations in the average distance from point-pair partners to their nearest atoms, see the color coding in Fig.~\ref{fig:Pairs_of_grid_point_with_various_molecule}, but generally these differences are small and do not correlate with any other features of the $\Delta\EX$ or WSS($\hat{T}_2$) distributions.

\begin{table}
\centering
\caption[short]{The average and standard deviation of $R_0$ and $\sigma$ obtained by fitting a $t$-distribution to the correlation features obtained from $|\Delta\EX|/E$, accumulated across all calculation methods and basis sets (excluding cc-pVDZ). $R_0$ and $\sigma$ values are reported in a.u. Standard deviations for varying basis set and method for each molecule are reported in the main table, with the mean and standard deviation over reported molecular means given at the bottom of the table.} \label{table:pairs_of_grid_points_dependency_to_molecules_R0_and_R0_width}
\begin{tabular}{ p{2cm} *{8}{p{1.2cm}<{\centering}}}
\toprule
  & \multicolumn{4}{c}{$R_0$ (a.u.)} & \multicolumn{4}{c}{$\sigma$ (a.u.)} \\ 
  \cmidrule(r){2-5}
  \cmidrule(r){6-9}
  &\multicolumn{4}{c}{$\chi$} &
  \multicolumn{4}{c}{$\chi$} \\
\cmidrule(r){2-5}
\cmidrule(r){6-9}

Molecule &  3.0 & 4.0 & 5.0 & 6.0 & 3.0 & 4.0 & 5.0 & 6.0 \\
\midrule 
Adenine    & 1.14 $\pm$ 0.07 & 1.32 $\pm$ 0.15 & 1.53 $\pm$ 0.12 & 1.65 $\pm$ 0.21 & 1.33 $\pm$ 0.19 & 1.40 $\pm$ 0.21 & 1.47 $\pm$ 0.15 & 1.53 $\pm$ 0.14\\
AT-ST      & 1.16 $\pm$ 0.05 & 1.28 $\pm$ 0.13 & 1.50 $\pm$ 0.11 & 1.60 $\pm$ 0.18 & 1.30 $\pm$ 0.16 & 1.35 $\pm$ 0.20 & 1.39 $\pm$ 0.11 & 1.45 $\pm$ 0.11\\
AT-WC      & 1.25 $\pm$ 0.09 & 1.36 $\pm$ 0.15 & 1.57 $\pm$ 0.14 & 1.68 $\pm$ 0.21 & 1.28 $\pm$ 0.17 & 1.35 $\pm$ 0.18 & 1.39 $\pm$ 0.10 & 1.44 $\pm$ 0.11\\
Benzene    & 1.24 $\pm$ 0.06 & 1.45 $\pm$ 0.17 & 1.60 $\pm$ 0.24 & 1.63 $\pm$ 0.25 & 1.57 $\pm$ 0.20 & 1.70 $\pm$ 0.22 & 1.80 $\pm$ 0.17 & 1.92 $\pm$ 0.19\\
PFBA       & 1.14 $\pm$ 0.13 & 1.32 $\pm$ 0.16 & 1.36 $\pm$ 0.22 & 1.42 $\pm$ 0.27 & 1.28 $\pm$ 0.12 & 1.27 $\pm$ 0.16 & 1.31 $\pm$ 0.15 & 1.34 $\pm$ 0.14\\
Ph-Criegee & 1.20 $\pm$ 0.08 & 1.36 $\pm$ 0.15 & 1.57 $\pm$ 0.24 & 1.65 $\pm$ 0.31 & 1.45 $\pm$ 0.17 & 1.53 $\pm$ 0.19 & 1.61 $\pm$ 0.13 & 1.71 $\pm$ 0.12\\
Thymine    & 1.26 $\pm$ 0.10 & 1.41 $\pm$ 0.10 & 1.60 $\pm$ 0.18 & 1.67 $\pm$ 0.31 & 1.34 $\pm$ 0.13 & 1.41 $\pm$ 0.16 & 1.46 $\pm$ 0.10 & 1.52 $\pm$ 0.11\\
Hexane     & 1.26 $\pm$ 0.15 & 1.49 $\pm$ 0.16 & 1.61 $\pm$ 0.13 & 1.69 $\pm$ 0.09 & 1.51 $\pm$ 0.20 & 1.60 $\pm$ 0.13 & 1.72 $\pm$ 0.10 & 1.83 $\pm$ 0.15\\
Octane     & 1.26 $\pm$ 0.17 & 1.45 $\pm$ 0.16 & 1.55 $\pm$ 0.14 & 1.64 $\pm$ 0.09 & 1.46 $\pm$ 0.21 & 1.55 $\pm$ 0.17 & 1.68 $\pm$ 0.14 & 1.78 $\pm$ 0.15\\
Decane     & 1.32 $\pm$ 0.10 & 1.53 $\pm$ 0.11 & 1.67 $\pm$ 0.11 & 1.74 $\pm$ 0.09 & 1.41 $\pm$ 0.18 & 1.48 $\pm$ 0.13 & 1.57 $\pm$ 0.10 & 1.67 $\pm$ 0.13\\ 
\midrule 
Mean & 1.22 & 1.40 & 1.56 &  1.64 &  1.39  & 1.47 & 1.54  & 1.62 \\
Std. Dev. & 0.06 & 0.08 & 0.08 &  0.09 &  0.10  & 0.13 & 0.16  & 0.19 \\
\bottomrule
\end{tabular} 
\end{table}


\subsubsection{Trends Across Wavefunctions}\label{sub:sub:pair_wavefunctions}

\begin{figure}
     \centering
     \includegraphics[width = \textwidth]{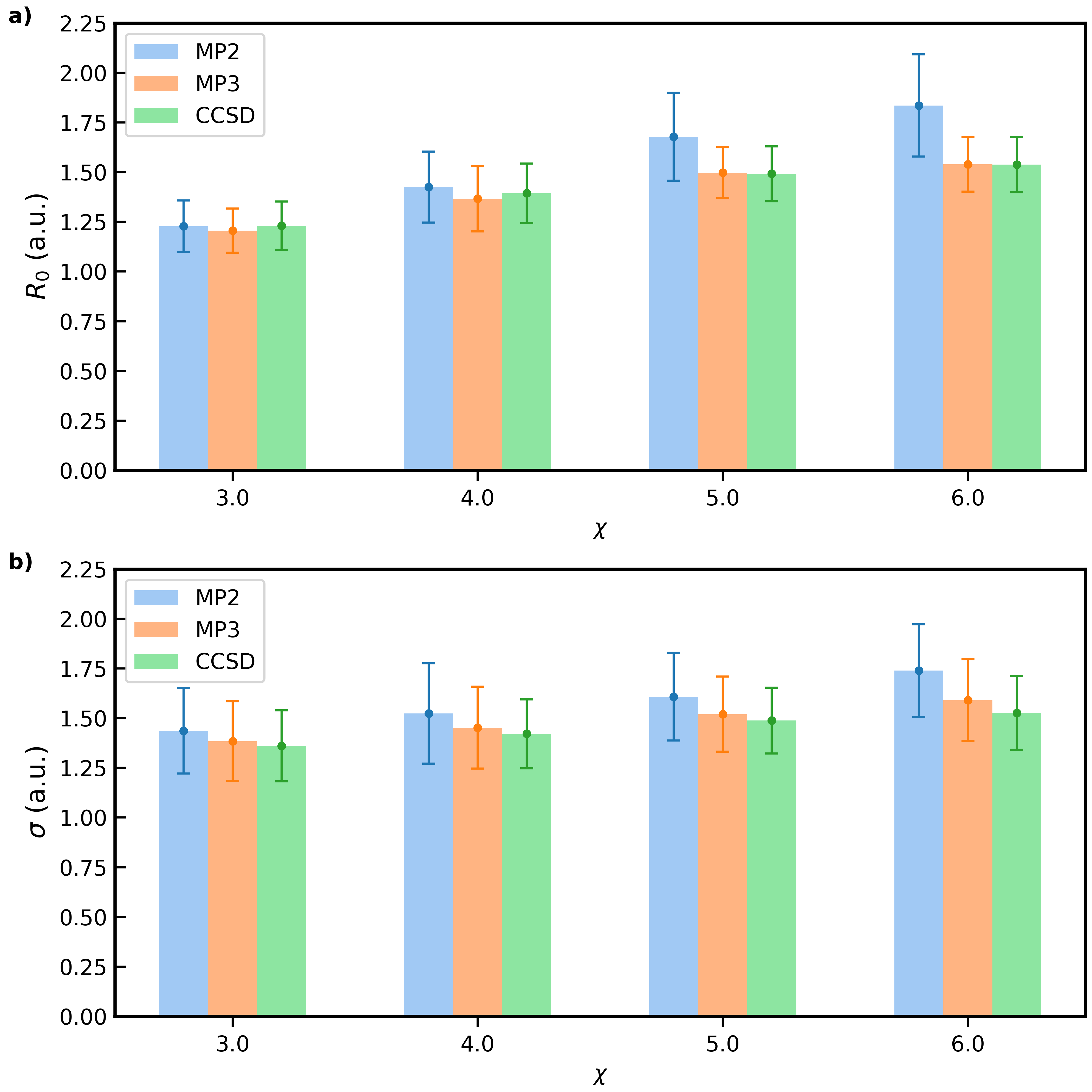}
     \caption{The average and standard deviation of \textbf{a)} $R_0$ and \textbf{b)} $\sigma$ of correlation features obtained from the fit of $|\Delta\EX|/E$ to a $t$-distribution for MP2, MP3, and CCSD at various values of $\chi$. Points indicate averages across all molecules and basis sets (excluding cc-pVDZ), with standard deviations indicated by whiskers.}     \label{fig:Pairs_of_grid_points_optimal_distance_with_various_methods}
\end{figure}

The correlation feature displays some slight dependence on the particular correlated wavefunction that LS-THC is approximating. Fig.~\ref{fig:Pairs_of_grid_points_optimal_distance_with_various_methods} and Table~\ref{table:pairs_of_grid_points_dependency_to_methods_R0_and_R0_width} display the $R_0$ of the $t$-distribution fit averaged across all molecules and basis sets (excluding cc-pVDZ) as a function of $\chi$ and correlated wavefunction. MP2, the least correlated of these wavefunctions, displays slightly more sensitivity to $\chi$ than MP3 and CCSD, while MP3 and CCSD exhibit very nearly identical profiles. To some extent this may not be very surprising: the MP2 wavefunction is well known to behave qualitatively differently from more correlated wavefunctions when exposed to factorization schemes like rank-reduction\cite{Parrish-JCP-2019,Hohenstein-JCP-2022}. Put simply, MP2 has ``less going on'' (quite literally, for example when considering diagrammatic contributions to the amplitudes) and thus the LS-THC single-point space does a qualitatively better job of describing the wavefunction. As discussed in the previous section, more complete single-point spaces push the correlation feature peak location to longer inter-pair distances with increasing $\chi$. As MP2 converges more quickly with regard to the single point grid than MP3 or CCSD (see Fig.~\ref{fig:partial_energy_methods}, this manifests as a shift to longer $R_0$ at earlier $\chi$. Counter to this increase in $R_0$, more correlated wavefunctions correspond to smaller correlation feature widths. As with $R_0$, this difference is most pronounced in the change from MP2 to MP3, while MP3 and CCSD $\sigma$ are more similar.

\begin{table}
\centering
\caption[short]{The average and standard deviation of $R_0$ and $\sigma$ of $|\Delta\EX|/E$ correlation features fit to a $t$-distribution at various $\chi$ for MP2, MP3, and CCSD wavefunctions. Values are accumulated across all molecules and basis sets (excluding cc-pVDZ).} \label{table:pairs_of_grid_points_dependency_to_methods_R0_and_R0_width}
\begin{tabular}{ p{3cm} *{8}{p{1.2cm}<{\centering}}}
\toprule
  &\multicolumn{4}{c}{$R_0$ (a.u.)} &
  \multicolumn{4}{c}{$\sigma$ (a.u.)} \\
\cmidrule(r){2-5}
\cmidrule(r){6-9}

\multicolumn{1}{r}{$\chi =$} &  3.0 & 4.0 & 5.0 & 6.0 & 3.0 & 4.0 & 5.0 & 6.0 \\
\midrule 
MP2  & 1.23 $\pm$ 0.13 & 1.43 $\pm$ 0.18 & 1.68 $\pm$ 0.22 & 1.84 $\pm$ 0.26 & 1.44 $\pm$ 0.22 & 1.52 $\pm$ 0.25 & 1.61 $\pm$ 0.22 & 1.74 $\pm$ 0.23 \\
MP3  & 1.21 $\pm$ 0.11 & 1.37 $\pm$ 0.16 & 1.50 $\pm$ 0.13 & 1.54 $\pm$ 0.14 & 1.38 $\pm$ 0.20 & 1.45 $\pm$ 0.21 & 1.52 $\pm$ 0.19 & 1.59 $\pm$ 0.21 \\
CCSD & 1.23 $\pm$ 0.12 & 1.39 $\pm$ 0.15 & 1.49 $\pm$ 0.14 & 1.54 $\pm$ 0.14 & 1.36 $\pm$ 0.18 & 1.42 $\pm$ 0.17 & 1.49 $\pm$ 0.17 & 1.53 $\pm$ 0.19 \\
\midrule 
Mean & 1.22 & 1.40 & 1.56 & 1.64	& 1.39 & 1.47 & 1.54 & 1.62 \\
Std. Dev. & 0.01 & 0.03 & 0.11 & 0.17	& 0.04 & 0.05 & 0.06 & 0.11 \\
\bottomrule
\end{tabular} 
\end{table}


\subsubsection{Trends Across Basis Sets}\label{sub:sub:pair_basis}

\begin{figure}[t]
     \centering
     \includegraphics[width = \textwidth]{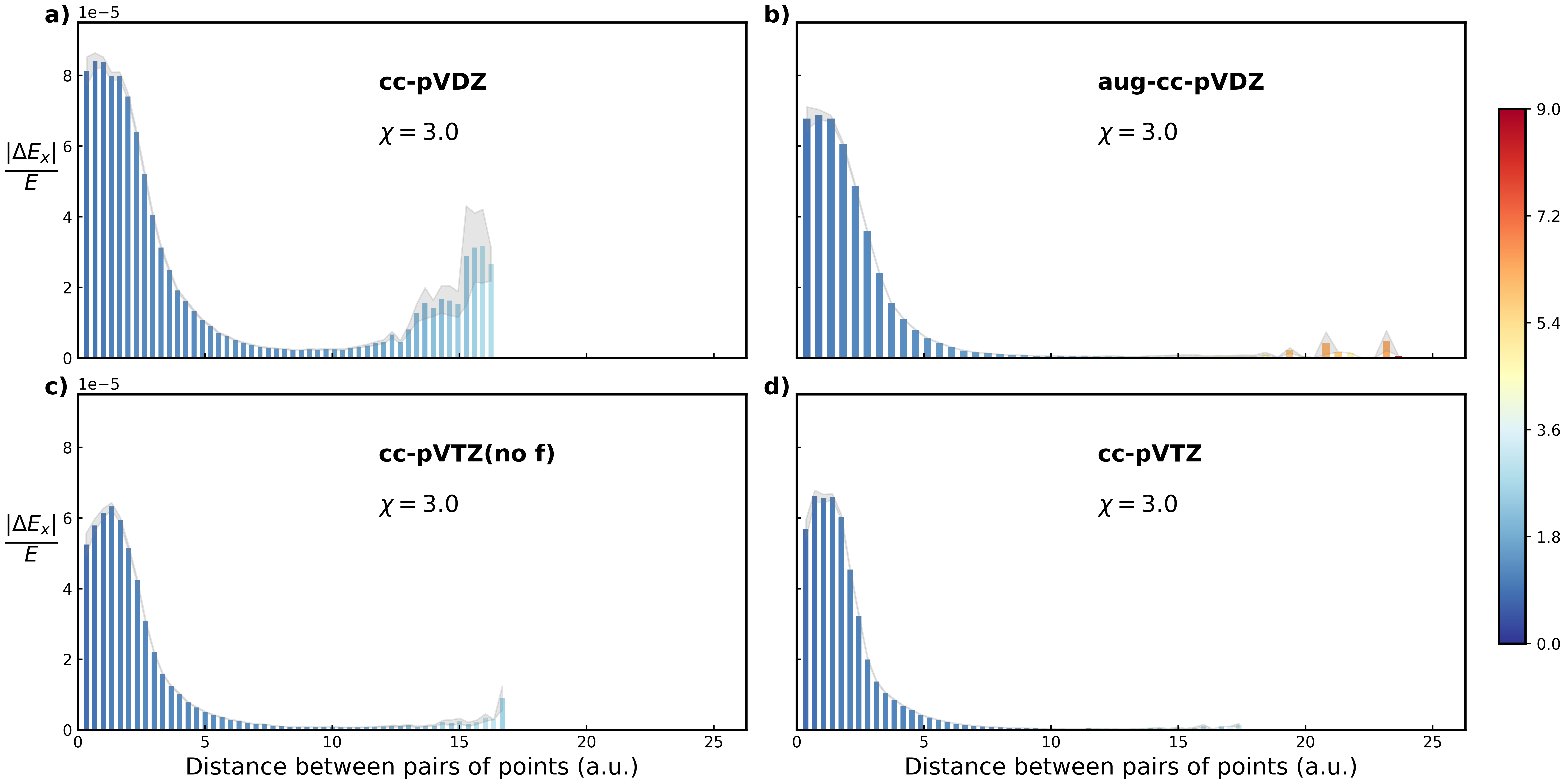}
     \caption{Spatial histograms for adenine at the CCSD level and with various basis sets. In each case $\chi = 3.0$, although the differing size of the orbital basis leads to differently sized grids in each case. See Fig.~\ref{fig:pairs_of_grid_points_example} for details.}
     \label{fig:Pairs_of_grid_points_basis_sets_with_fixed_chi}
 \end{figure}

\begin{figure}
     \centering
     \includegraphics[width = \textwidth]{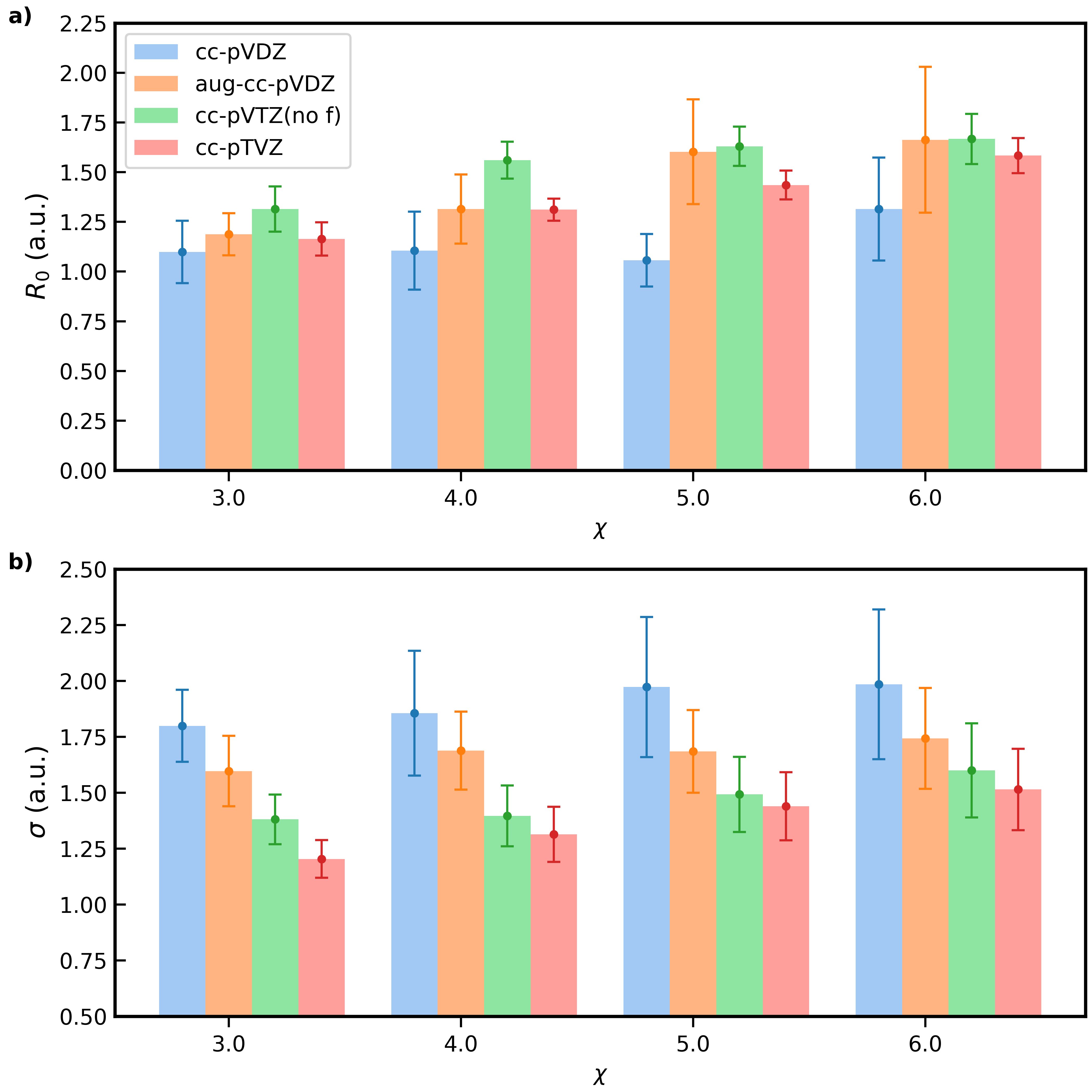}
     \caption{The average and standard deviation of \textbf{a)} $R_0$ and \textbf{b)} $\sigma$ of correlation features obtained from the fit of $|\Delta\EX|/E$ to a $t$-distribution for cc-pVDZ, aug-cc-pVDZ, cc-pVTZ(no~f), and cc-pVTZ at various values of $\chi$. Points indicate averages across all molecules and correlation methods, with standard deviations indicated by whiskers.}\label{fig:Pairs_of_grid_points_optimal_distance_with_various_basis_sets}
 \end{figure}

Fig.~\ref{fig:Pairs_of_grid_points_basis_sets_with_fixed_chi} displays some properties of the basis set dependence of the correlation feature discussed here. Variation across basis set is slightly more complicated than variation across molecules and wavefunctions, as the number of basis functions, and therefore the definition of $\chi$, is directly related to the basis set in question. Simply put, the ``quality'' of the single-point LS-THC basis varies more at the same $\chi$ when the basis set is changed than when the molecule or wavefunction is changed. Of note, the cc-pVDZ basis set features systematically shorter peak centers in the fit of the correlation feature, while cc-pVTZ(no~f) features the longest though by a much smaller margin. This is particularly clear in Fig.~\ref{fig:Pairs_of_grid_points_optimal_distance_with_various_basis_sets} and Table~\ref{table:pairs_of_grid_points_dependency_to_basis_sets_R0_and_R0_width}, which display $R_0$ as a function of $\chi$ for the various basis sets here, averaged over molecules and wavefunctions. Additionally, cc-pVDZ and aug-cc-pVDZ basis sets feature larger standard deviations than cc-pVTZ(no~f) and cc-pVTZ, indicating that the former two basis sets result in greater variation between species and wavefunction than the latter two. Finally, while the correlation feature for aug-cc-pVDZ, cc-pVTZ(no~f), and cc-pVTZ shifts systematically to longer inter-pair distances, the average peak position of cc-pVDZ shifts to shorter inter-pair distances between $\chi = 3$ and $\chi = 5$, before climbing again for $\chi = 6$. We suspect this inconsistent behavior is due to the very small number of basis functions present in the cc-pVDZ basis set, which, when used as the denominator in $\chi$, results in relatively fewer grid points per basis function to capture features of the molecules. 

Fig.~\ref{fig:Pairs_of_grid_points_basis_sets_with_fixed_chi} also highlights our observation that, while the convergence of the exchange-like energy in the cc-pVDZ basis with $\chi$ is not uniformly faster or slower than in other basis sets (data not shown, raw data is available in the SI), both long-range and short-range points are incorporated into the this grid at relatively larger $\chi$ compared to other basis sets. For example, one can see that the long(er)-range points clearly still contribute in Fig.~\ref{fig:Pairs_of_grid_points_basis_sets_with_fixed_chi}a but not in other basis sets at $\chi = 3$. We also compared spatial histograms across basis sets with a constant grid size rather than constant $\chi$ and observed the opposite effect (cc-pVDZ was more ``converged'': diminished short- and long-range contributions with more decay of the correlation feature). This suggests a need for a slightly larger grid size for cc-pVDZ relative to other basis sets for the same quality. The same situation is encountered in the density fitting approximation, where relatively larger auxiliary grids (for example, as measured as a multiple of the orbital basis size) are required for double-$\zeta$ quality basis sets.\cite{Weigend-JCP-2002}

\begin{table}
\centering
\caption[short]{The average and standard deviation of $R_0$ and $\sigma$ of $|\Delta\EX|/E$ correlation features fit to a $t$-distribution at various $\chi$ for cc-pVDZ, aug-cc-pVDZ, cc-pVTZ(no~f), and cc-pVTZ basis sets averaged across molecules and correlation methods.} \label{table:pairs_of_grid_points_dependency_to_basis_sets_R0_and_R0_width}
\begin{tabular}{ p{3cm} *{8}{p{1.2cm}<{\centering}}}
\toprule
  &\multicolumn{4}{c}{$R_0$ (a.u.)} &
  \multicolumn{4}{c}{$\sigma$ (a.u.)} \\
\cmidrule(r){2-5}
\cmidrule(r){6-9}

\multicolumn{1}{r}{$\chi =$} &  3.0 & 4.0 & 5.0 & 6.0 & 3.0 & 4.0 & 5.0 & 6.0 \\
\midrule 
cc-pVDZ       & 1.10 $\pm$ 0.16 & 1.11 $\pm$ 0.20 & 1.06 $\pm$ 0.13 & 1.31 $\pm$ 0.26  & 1.80 $\pm$ 0.16 & 1.86 $\pm$ 0.28 & 1.97 $\pm$ 0.31 & 1.98 $\pm$ 0.34 \\
aug-cc-pVDZ   & 1.19 $\pm$ 0.11 & 1.31 $\pm$ 0.17 & 1.60 $\pm$ 0.26 & 1.66 $\pm$ 0.37  & 1.60 $\pm$ 0.16 & 1.69 $\pm$ 0.17 & 1.69 $\pm$ 0.19 & 1.74 $\pm$ 0.23\\
cc-pVTZ(no~f) & 1.31 $\pm$ 0.11 & 1.56 $\pm$ 0.09 & 1.63 $\pm$ 0.10 & 1.67 $\pm$ 0.13  & 1.38 $\pm$ 0.11 & 1.40 $\pm$ 0.14 & 1.49 $\pm$ 0.17 & 1.60 $\pm$ 0.21\\
cc-pVTZ       & 1.16 $\pm$ 0.08 & 1.31 $\pm$ 0.06 & 1.44 $\pm$ 0.07 & 1.58 $\pm$ 0.09  & 1.20 $\pm$ 0.08 & 1.31 $\pm$ 0.12 & 1.44 $\pm$ 0.15 & 1.51 $\pm$ 0.18\\
\midrule 
Mean & 1.20	& 1.33 & 1.43 & 1.55 & 1.59	& 1.65 & 1.72 &1.78 \\ 
Std. Dev. & 0.11	& 0.23 & 0.32 & 0.20 & 0.21	& 0.23 & 0.24 &0.19 \\ 
\bottomrule
\end{tabular} 
\end{table}



In contrast to the peak center, the peak width ($\sigma$) systematically shrinks as the size of the basis increases. It should be noted that this does not indicate that these basis sets are better represented by LS-THC, rather it indicates there are pairs of points with a tighter spread of inter-pair distances that are neglected by the LS-THC factorization at a given $\chi$. The interaction of the size of the starting single-point grid with the quality of the orbital basis is a contributor here, for example in the higher angular and radial resolution afforded by larger basis sets.

\section{Discussion}\label{sub:sub:pair_conclusion}

The above results provide a strong indication that the poor performance of LS-THC in the representation of $\EX$ and the $\hat{T}_2$ amplitudes arises as a result of the missing contribution from pair points (in the sense of \eqref{eq:t2-quadrature}) to the representation of the doubles amplitudes, and that the important pair points follow a consistent spatial distribution. The following properties of the amplitudes and pair point distribution appear to be important, and are worth enumerating:
\begin{enumerate}[a)]
    \item{The error in the LS-THC representation of $\hat{T}_2$ manifests primarily in the exchange-like terms of electron correlation and in the WSS of $\hat{T}_2$ amplitudes, but it does not directly affect the Coulomb-like energy term. Note that this analysis presupposes an MP2- or CCSD-like energy functional---however insufficiency in the representation of the doubles amplitudes can manifest even in other energy functionals, such as in the MP3b formulation which uses $\Delta E_\text{MP3} = \langle 0|\hat{T}_2^{[1]\dagger} \hat{V} \hat{T}_2^{[1]}|0\rangle$\cite{Matthews-JCP-2021}.}
    \item{The additional contribution (energy or WSS) from pair points follows an exponentially-decaying distribution centered on inter-pair distances of one to two a.u. with a width of one to two a.u.}
    \item{On average, the pair points that contribute most strongly to $\EX$ and $\hat{T}_2$ are located relatively close to the atoms in the system, within about 2 a.u.}
    \item{As the size of the initial single point grid increases, the center of the correlation feature shifts to pair points with slightly longer inter-pair distances, while the width of the correlation feature increases slightly as well.}
    \item{The shape of this ``correlation feature'' for ground state wavefunctions is largely invariant to molecular species, correlation method, and single-particle basis sets.}
\end{enumerate}
The decay of importance of pair points with increasing inter-pair distance is not particularly surprising given the inherently short-range nature of electronic correlation, such as postulated by Kohn\cite{Kohn-PRL-1996} and of critical importance to methods relying on orbital localization for scaling reduction.\cite{Saebo-CPL-1985, Haser-TCA-1993, Ayala-JCP-1999,Song-JCP-2016, Werner-JCP-2003, Hattig-JCP-2012, Werner-JCTC-2015, Neese-JCP-2009, Hansen-JCP-2011, Riplinger-JCP-2013, Guo-JCP-2018,  Kitaura-CPL-1999, Li-MP-2016, Flocke-JCP-2004, Jin-JPCA-2019, Rolik-JCP-2011, Nagy-JCTC-2019, Eriksen-JCTC-2015} What is perhaps more surprising is the apparent stability of the observed spatial distribution across molecular systems and correlation methods. This hints at a universal and more fundamental basis for the emergence of the observed features, as well as a potentially transferable and robust method for correcting LS-THC errors in complex wavefunction methods.

One potential analysis can be made based on a picture of dispersion-dominated correlation interactions. In this picture, the $\hat{t}$ kernel of \eqref{eq:t2-quadrature} represents essentially a classical dipole-dipole interaction, while the integrals over orbital pairs represent the distribution of charge fluctuations. Discretization via quadrature gives a concrete realization of this distribution as the sum over a large number of instantaneous classical dipoles, weighted by both the quadrature (essentially, point volume) and something like a transition density. Thus, the observed spatial distribution would indicate a distribution of the length of the fluctuating dipoles. As Kohn noted, such fluctuations are due to quantum interference, and hence should decay on a length scale of the order of the electron de Broglie wavelength. Taking a typical valence electron kinetic energy for an organic molecule (e.g. calculated from $\langle \phi_p |-\frac{1}{2}\nabla^2|\phi_p\rangle$) of $\sim 1.2$ a.u., this gives a de Broglie wavelength of $\sim 4$ a.u. This compares well with the observed decay of the pair point inter-pair distance distribution, although of course such a comparison is highly qualitative. The behavior of the inter-pair distance distribution at short inter-pair lengths has multiple potential explanations. For example, the short-range value of the transition density is driven by overlap of the occupied and virtual orbitals, and thus the difference in general diffuseness of occupied and virtual orbitals could decrease the contribution in this range. Likewise, a pair point, at short inter-pair distance, becomes linearly dependent with the single point representing the pair center (or equivalently, either point in the pair itself). If these points are already included in the single point grid then the pair point contribution goes to zero.

Regardless of the true form of the correlation kernel, this work provides some important guidelines for the future development of LS-THC. First, it may be prudent to consider LS-THC grid point basis as two complementary components: a single point basis and a corresponding pair-point basis. Given the rapid convergence of $\EC$ as a function of $\chi$ (and the excellent representation of the ERIs reported in previous literature\cite{Parrish-JCP-2012,lee-JCTC-2020,Matthews-JCP-2020,Matthews-JCP-2021}), it is likely that this single point space need not be very large at all, and the current selection based upon Cholesky pivoting is entirely satisfactory for this purpose\cite{Matthews-JCTC-2020}. The pair-point space could then be selected either with a cut-off radius between pairs of points, or with a weighting function that follows the general shape of the correlation feature discovered here. Fortuitously, the majority of the character of this pair-point kernel appears to depend on less than 15\% of the total possible space of pair-points and is an inherently local phenomenon. A better implementation would depend upon the nature of the pair-point kernel itself (are the pairs of points atomic centered, located on neighboring atom pairs, etc.), and we expect that further study of this phenomena will result in further improved LS-THC treatments of ground state electron correlation.

\section{Conclusions}
In this work we have examined in some detail the grid point-dependence of LS-THC methods in the description of ground-state valence electron correlation. We demonstrate for the first time that the Coulomb-like ($\EC$) and exchange-like ($\EX$) valence correlation energies of MP2, MP3, and CCSD, with the cc-pVDZ, aug-cc-pVDZ, cc-pVTZ(no f), and cc-pVTZ basis sets, all follow the same bi-exponential dependence on the ratio $\chi$ of the number of LS-THC grid points to the number of basis functions. The poor performance of LS-THC in the description of exchange correlation energy arises from the fact that the LS-THC $\EX$ is still missing 5\% or more of the total exchange correlation energy by the time the Coulomb correlation has converged to within 0.1\% of its final value. 

Following the recent analysis of Ref.~\citenum{Mardirossian-JCP-2018}, we have developed a novel set of theoretical tools to examine the importance of pairs of spatial grid points in the LS-THC representation of the correlation energy (via ``energy partitioning'') and the correlated wavefunction (via ``weighted subspace score'', or WSS). Using these methods, we were able to efficiently examine the influence of 50,000 randomly selected pairs of grid points for various values of $\chi$ on a suite of ten molecules with all permutations of aforementioned correlation methods and basis sets. Following this analysis, we examine, for the first time, the spatial distribution of pair points in the LS-THC description of valence correlation energy and wavefunctions. 

Our principal finding is the spatial signature of a ``correlation feature'' in the dependence of $\Delta\EX$ and WSS($\hat{T}_2$) on the inter-pair distance of pair points and their proximity to atomic centers. Importantly, this feature emerges at a range of $\chi$ values corresponding to the transition between ``fast'' and ``slow'' convergence in the single point grid, and then decays very slowly with further increases in $\chi$. We postulate that this correlation feature ultimately derives from a universal structure of the ``correlation kernel'' which defines the $\hat{T}_2$ amplitudes in a real-space interpretation, and which we believe to be responsible for the comparatively poor description of $\EX$ and $\hat{T}_2$ that LS-THC presents at reasonable values of $\chi$.  

We conclude by reiterating some important properties of the observed correlation feature, primarily with regards to its spatial distribution but also with regard to its remarkable consistency across basis set, correlation method, and molecular species. This correlation feature is roughly Gaussian in shape and centered on pairs of points with inter-pair distances of $R_0\approx1$--2 a.u. and of width $\sigma \approx 1$--2 a.u. It also appears to favor pairs of grid points that are, on average, relatively close to the atoms of the molecule in question, generally with an average pair-partner distance to the nearest atom of less than $2$ a.u. With increasing single point grid coverage (measured by $\chi$) the peak of this feature shifts towards pairs of points with slightly longer inter-pair distances, and the width of this peak correspondingly increases. More correlated wavefunctions favor slightly shorter inter-pair distances with a tighter distribution. Basis set dependence of the correlation peak's center does not show any clear trends, but the width of the feature clearly shrinks as the size of the size of the basis set increases.

Future studies will investigate how to exploit these spatial signatures to improve both the accuracy and efficiency of LS-THC, and will make this already promising technique an even more viable tool in the accurate simulation of large molecules.

\begin{acknowledgement}

This study was supported by Department of Energy (Grant No. DE-SC0022893). J.H.T. acknowledges support from the SMU Moody School of Graduate and Advanced Studies. Computational resources for this research were provided by SMU’s O’Donnell Data Science and Research Computing Institute.

\end{acknowledgement}

\begin{suppinfo}
See the support material for the optimized geometry of all molecules used in this work, the raw partial energy profile data for exchange and Coulomb energy(.csv), the raw pairs of grid points energy and weighted subspace score data (.csv), the scripts (.py) for plotting the figures and partial energy fitting and pairs of grid points fitting.

\end{suppinfo}

\section{Appendix: Pairs of Grid Points Energy Analysis Derivation}

\setcounter{equation}{0}
\renewcommand{\theequation}{A\arabic{equation}}

Here, we present the detailed derivation of the working equations for the pair grid point energy analysis. Using $\EC$ as an example, we reiterate,
\begin{align}
\EC^\prime = \mathbf{g}\mathbf{Y^\prime}(\mathbf{S^\prime})^{-1}(\mathbf{Y^\prime})^{T}\mathbf{t}\mathbf{Y^\prime}(\mathbf{S^\prime})^{-1}(\mathbf{Y^\prime})^{T})\label{eq:energy-appendix}
\end{align}
using the block-matrix form of $\mathbf{Y^\prime}$,

\begin{align}
\mathbf{Y'} &=
\begin{bmatrix}
 \mathbf{Y} & \mathbf{\Delta Y}
\end{bmatrix}
\end{align}
We can also write the inverse of the metric matrix in block form,
\begin{align}
(\mathbf{S^\prime})^{-1} &=
\begin{bmatrix}
 \mathbf{A} &   \mathbf{B}\\
  \mathbf{B}^T & \mathbf{D}
\end{bmatrix}^{-1}
\end{align}
where $\mathbf{A} =\mathbf{S} $, $\mathbf{B} = \mathbf{Y}^T\mathbf{\Delta Y}$, $\mathbf{D} = \mathbf{\Delta Y}^T \mathbf{\Delta Y}$.
The block matrix inverse yields,
\begin{align}
(\mathbf{S^\prime})^{-1} &=
\begin{bmatrix}
    \mathbf{A}^{-1} +  \mathbf{A}^{-1}\mathbf{B} \mu \mathbf{B}^{T}\mathbf{A}^{-1} & -  \mathbf{A}^{-1} \mathbf{B}\mu \\
    - \mu  \mathbf{B}^{T}\mathbf{A}^{-1} &  \mu 
\end{bmatrix}
\end{align} \\
where $\mu =  (\mathbf{D} - \mathbf{B}^{T}\mathbf{A}^{-1}\mathbf{B})^{-1} = (\mathbf{\Delta Y}^T \mathbf{\Delta Y} - \mathbf{\Delta Y}^T\mathbf{Y} \mathbf{S}^{-1}\mathbf{Y}^T\mathbf{\Delta Y})^{-1}$. When adding only a single pair grid point, this is a scalar quantity and is easily inverted.
Using this, we further find, 
\begin{align}
\mathbf{Y^\prime}(\mathbf{S^\prime})^{-1}(\mathbf{Y^\prime})^{T} & = 
\begin{bmatrix}
    \mathbf{Y} & \mathbf{\Delta Y}
\end{bmatrix}
\begin{bmatrix}
    \mathbf{A}^{-1} +  \mathbf{A}^{-1}\mathbf{B} \mu \mathbf{B}^{T}A^{-1} & -  A^{-1} \mathbf{B}\mu \\
    - \mu  \mathbf{B}^{T}\mathbf{A}^{-1} &  \mu 
\end{bmatrix}
\begin{bmatrix}
    \mathbf{Y} & \mathbf{\Delta Y }
\end{bmatrix} ^{T}
\\
& = \mathbf{Y}\mathbf{S}^{-1}\mathbf{Y}^{T} +  \mathbf{C} \\
\mathbf{C} &= (\mathbf{Y}\mathbf{S}^{-1}\mathbf{Y}^T - \mathbf{I})\mathbf{\Delta Y} \mu \mathbf{\Delta Y}^{T} (\mathbf{Y}\mathbf{S}^{-1}\mathbf{Y}^T - \mathbf{I})
\end{align}
In practice, we perform a pivoted Cholesky decomposition of the single point metric matrix $\mathbf{S}$,\cite{Matthews-JCP-2020} which allows for some simplification,
\begin{align}
\mathbf{S} &= \mathbf{L}\mathbf{L}^T \\
\mathbf{WL}^T &= \mathbf{Y} \\
\mathbf{Y}\mathbf{S}^{-1}\mathbf{Y}^{T} &= \mathbf{WW}^T \label{eq:YSY-final} \\
\mathbf{z} &= \mathbf{W}^T \mathbf{\Delta Y} \\
\mathbf{C} &= (\mathbf{Wz} - \mathbf{\Delta Y})\mu(\mathbf{Wz} - \mathbf{\Delta Y})^T \nonumber \\
          &= \mathbf{d} \mu \mathbf{d}^T \label{eq:C-final}
\end{align}
Substituting \eqref{eq:YSY-final} and \eqref{eq:C-final} into \eqref{eq:energy-appendix}, we find,
\begin{align}
    \EC^\prime &= 2\Tr \left[ \mathbf{g} \mathbf{WW}^T \mathbf{t} \mathbf{WW}^T \right] + 4\Tr \left[ \mathbf{g} \mathbf{WW}^T \mathbf{t} \mathbf{d}\mu\mathbf{d}^T \right] + 2\Tr \left[ \mathbf{g} \mathbf{d}\mu\mathbf{d}^T \mathbf{t} \mathbf{d}\mu\mathbf{d}^T \right] \nonumber \\
    &= 2\Tr \left[ \mathbf{W}^T \mathbf{E} \mathbf{W} \right] + 4\Tr \left[ \mu\mathbf{d}^T \mathbf{E} \mathbf{d} \right] + 2\Tr \left[ (\mu\mathbf{d}^T \mathbf{g} \mathbf{d} )( \mu\mathbf{d}^T \mathbf{t} \mathbf{d} ) \right] \label{eq-factorized} \\
    \mathbf{E} &= \frac{1}{2}(\mathbf{g} \mathbf{WW}^T \mathbf{t} + \mathbf{t} \mathbf{WW}^T \mathbf{g})
\end{align}
When computing $\Delta \EC$ (the second and third term of \eqref{eq-factorized}) separately for a large set of pair grid points, we can prepare the starting energy $\EC$ (first term of \eqref{eq-factorized}) and the matrix $\mathbf{E}$ in $\mathcal{O}(n_o^2 n_v^2 n_g)$ time. Computing the energy changes for $n_2$ pair grid points then requires only $\mathcal{O}(n_o^2 n_v^2 n_2)$ additional time and with efficient level-3 BLAS operations.

\bibliography{achemso-demo}

\end{document}